# Social Distancing Induced Coronavirus Optimization Algorithm (COVO): Application to Multimodal Function Optimization and Noise Removal

Om Ramakisan Varma[1]
Research Scholar, CSE Department
NITTTR, Chandigarh, India
om.cse20@nitttrchd.ac.in

Mala Kalra[2]
Assistant Professor, CSE Department
NITTTR, Chandigarh, India
varma.om@gmail.com

***Abstract*— The metaheuristic optimization technique attained more awareness for handling complex optimization problems. Over the last few years, numerous optimization techniques have been developed that are inspired by natural phenomena. Recently, the propagation of the new COVID-19 implied a burden on the public health system to suffer several deaths. Vaccination, masks, and social distancing are the major steps taken to minimize the spread of the deadly COVID-19 virus. Considering the social distance to combat the coronavirus epidemic, a novel bio-inspired metaheuristic optimization model is proposed in this work, and it is termed as Social Distancing Induced Coronavirus Optimization Algorithm (COVO). The pace of propagation of the coronavirus can indeed be slowed by maintaining social distance. Thirteen benchmark functions are used to evaluate the COVO performance for discrete, continuous, and complex problems, and the COVO model performance is compared with other well-known optimization algorithms. The main motive of COVO optimization is to obtain a global solution to various applications by solving complex problems with faster convergence. The error in the fitness function for the COVO model is 1.23×10−18, which is significantly lower than the error values achieved by other existing models: EHO at 4.38×10−8, SSA at 10.12, SSO at 4.05×10−10, SFO at 26.5, BOA at 4.20×10−9, BWO at 3.58×10−5, SMO at 1.04, CVOA 2.95×10−8, SRO 2.40×10−8, and GBRUN 13.27. At last, the validated results depict that the proposed COVO optimization has a reasonable and acceptable performance.***

**Nomenclature**

| **Abbreviation** | **Description** |
|---|---|
| ABC | Artificial Bee Colony |
| ACO | Ant Colony Optimization |
| BA | Bat Algorithm |
| BOA | Butterfly Optimization Algorithm |
| BWO | Black Widow Optimization |
| BXOA | Botox Optimization Algorithm |
| C-19BOA | COVID-19 Based Optimization Algorithm |
| COVO | Social Distancing Induced Coronavirus Optimization Algorithm |
| CS | Cuckoo Search |
| CVA | COVID-19 Optimizer Algorithm |
| CVO | Coronavirus Optimization |
| CVOA | Coronavirus Optimization Algorithm |
| CVSO | Corona Virus Search Optimizer |
| EA | Evolutionary Algorithms |
| CHIO | Coronavirus Herd Immunity Optimizer |
| EHO | Elephant Herding Optimization |
| FA | Firefly Algorithm |
| GA | Genetic Algorithm |
| GBRUN | Gradient Search-Based Binary Runge Kutta Optimizer |
| HC | Hill Climbing |
| HSA | Harmony Search Algorithm |
| ICA | Independent Component Analysis |
| MCHIAO | Modified Coronavirus Herd Immunity Aquila Optimization |
| MOCOVIDOA | Multi-Objective Coronavirus Disease Optimization Algorithm |
| MPA | Marine Predator Algorithm |
| NFL | No Free Lunch |
| POA | Pufferfish Optimization Algorithm |
| PSO | Particle Swarm Optimization |
| SA | Simulated Annealing |
| SFO | Sailfish Optimizer |
| SMO | Spider Monkey Optimization |
| SSA | Sparrow Search Algorithm |
| SSO | Salp Swarm Optimization |
| SRO | Ship Rescue Optimization |
| VBA | Virtual Bee Algorithm |
| VOA | Virus Optimization Algorithm |
| WHO | World Health Organization |

## I. INTRODUCTION

Engineering design optimization is complex due to non-linearity and strict design rules, where conventional algorithms often miss global optimality. Metaheuristic algorithms are increasingly popular for efficient design, simulating physical processes while adapting to unknown factors and resource limitations [1].

The optimization algorithm is an integral part of any optimization process [2]. An optimization algorithm simulates physical processes and is key to efficient, durable solutions, requiring multiple evaluations. Image processing operations in computer vision aim to reduce complexity or improve image quality [3].

Optimizers are essential for finding optimal solutions, as there are various optimization methods. Gradient-based and gradient-free algorithms are used [4], with gradient-based techniques like steepest descent and Gauss-Newton using derivative information, and the Nelder-Mead downhill simplex method using goal values [5].

Deterministic and stochastic algorithms are two types of algorithms. A deterministic algorithm operates in a mechanically deterministic manner without any randomness. If we start with the same beginning position, such an algorithm will arrive at the same end position. Deterministic algorithms like hill climbing (HC) and downhill simplex are notable examples [6]. Stochastic algorithms introduce randomness, leading to different outcomes even with the same starting point.

Randomization in algorithms, such as genetic algorithms, introduces unpredictability through methods like crossover and mutation rates. Heuristics use trial and error for problem-solving, while metaheuristics operate at a higher level, guiding these search strategies [6], [7].

The metaheuristic approach provides a flexible optimization model that enhances solutions through adaptive, probabilistic processes guided by variable adjustments until an optimal result is achieved [8]. Sophisticated metaheuristic algorithms explore and exploit the search space, leveraging accumulated knowledge to solve problems. They adapt to various optimization tasks, require no derivative knowledge, and can avoid local optima [9], [10].

Remarkably, most metaheuristic-based techniques are primarily derived from natural events and may be divided into four categories: evolutionary, swarm, physical, and human-based algorithms [11], [12]. Evolutionary algorithms (EA) simulate natural selection, selecting the fittest individuals for reproduction over generations. John Henry Holland introduced genetic algorithms (GA) in 1960, inspired by Darwin's concept of gradual evolution [13] [14]. The collaborative behaviors of animal swarms inspire swarm-based algorithms. Particle swarm optimization (PSO), mimicking bird swarming, is this type's most popular and widely used algorithm. [15]. Particles (solutions) navigate the search space to find the global best, noting the local best along the way. This approach includes various swarm-based optimizers, such as ant colony optimization (ACO) and artificial bee colony (ABC) [16]. Physical-based algorithms, such as simulated annealing (SA), are grounded in cosmic principles and mimic the thermodynamic cooling process of metal annealing [17]. Finally, human-based algorithms inspire humans' habits, livelihood, or perspectives. The harmony search algorithm (HSA) [18] is indeed a fundamental technique of this class, wherein a collection of JAZZ players rehearse their devices' tones unless they achieve a pleasant harmony (optimal solution). Firefly algorithm (FA) [19] and many others constitute additional, prominent human-based algorithms. There appears to be a multitude of nature-inspired algorithms that may have been utilised to solve various optimization issues effectively. However, an optimization technique cannot achieve good performance for all optimization problems, as per the "No Free Lunch (NFL)" hypothesis [20].

Many deterministic and heuristic optimization techniques struggle with non-linearity and multimodality, highlighting the need for new nature-inspired algorithms to address complex objective functions effectively. Recently, a human-based nature-inspired phenomenon has arisen since techniques like HSA [18] and $\beta$ -hill climbing ( $\beta$ HC) [21] produce positive findings, contrasting to specific other nature-inspired algorithms.

The COVID-19 epidemic has inspired the development of various metaheuristic algorithms, reflecting our dynamic modern environment. The CVA [22] draws inspiration from strategies aimed at containing epidemics and reducing infection rates. The CVA fails to address the spread of infection between countries, resulting in low algorithm convergence accuracy and premature convergence. CHIO has several drawbacks, such as complex mathematical formulations, inadequate real-time optimization, numerous control parameters, and a high convergence tolerance. CVSO [23] is inspired by the spread of the Coronavirus across different communities, focusing on its movement, prevalence, and death rate. Its limitation lies in the scope and size of these communities, which can significantly affect the virus's transmission and intensity.

Researchers are developing metaheuristics that mimic the spread patterns of the coronavirus. COVID-19, caused by the SARS-CoV-2 virus, spreads much faster than typical respiratory illnesses like the cold. Social distancing aims to reduce transmission and delay, decrease the epidemic peak, and relieve pressure on healthcare systems. Traditional optimization algorithms, such as genetic and particle swarm optimization, often struggle with complex, high-dimensional, and dynamic problems, facing issues like premature convergence and poor exploration of solution spaces. To address these challenges, there is an increasing demand for innovative metaheuristic algorithms that offer improved efficiency, adaptability, and convergence. This study introduces the Social Distancing Induced Coronavirus Optimization Algorithm (COVO), inspired by the concept of social distancing to reduce COVID-19 transmission. The significant contributions of this research work are as follows.

- Proposing a novel Social Distancing Induced Coronavirus Optimization Algorithm (COVO)
- The proposed COVO algorithm considers the social distancing parameter, which has been suggested as a significant solution for reducing the spreading rate of COVID-19.
- The proposed COVO algorithm is validated using 13 standard benchmark functions (F1-F13).
- Demonstrates COVO's effectiveness by comparing its performance with several well-known optimization algorithms, showing competitive results regarding convergence speed and solution accuracy.

- The COVO method is applied to a real-world signal noise removal problem, demonstrating its practical utility and effectiveness in enhancing signal quality through optimized noise reduction.

The rest of this paper is organized as follows: Section II addresses recent literature on metaheuristic algorithms. Section III details the proposed COVO algorithm. The results acquired with the proposed work are comprehensively discussed in Section IV. The paper is concluded in Section V.

## II. LITERATURE REVIEW

### A. *Related works*

Simulated Annealing (SA) [17] is one of the first metaheuristic algorithms, and it was essentially a modified version of the Metropolis-Hastings algorithm that was used in a different situation [24]. Genetic algorithms' three essential components are crossover, mutation, and fittest selection. Each solution is represented by a chromosome, a string of characters. A crossover of two-parent strings generates offspring by swapping parts or genes of the chromosomes.

Ant Colony Optimization (ACO), replicate ant foraging behavior, using pheromone concentrations as chemical message [25]. These algorithms excel in discrete optimization problems, but research is ongoing, and few approaches currently address the associated challenges.

Particle swarm optimization (PSO) is based on natural swarm behaviour like fish and bird schooling. In a quasi-stochastic way, this method explores the space fitness function by changing the particle (search agents) trajectories as piecewise routes generated by positional vectors. The Harmony Search Algorithm (HSA) [18] is a music-inspired algorithm developed with the help of a musician's description of the improvisation process.

Cuckoo Search (CS) [26] is used to find optimal solutions to complex problems by mimicking the process of cuckoos laying eggs and the host birds' efforts to identify and remove foreign eggs from their nests. According to recent research, CS has the potential to be considerably more efficient than PSO and GA. Tabu search [27], and artificial immune system [28] are other metaheuristic algorithms that are as popular and efficient.

Bee algorithms mimic different aspects of bee behavior, with forager bees optimizing nectar collection by exploring various food sources [8]. Pheromone concentrations can be more directly connected to goal functions in the virtual bee algorithm (VBA) invented by Xin-She Yang in 2005 [29]. D. Karaboga, on the other hand, invented the ABC optimization method in 2005 [30].

The bat algorithm (BA) [31] is a recent metaheuristic model inspired by the echolocation behaviour of microbats. Most bats use brief, frequency-modulated sounds spanning about an octave, while some rely on constant-frequency signals.

J C Bansal et al. [32] have developed Spider Monkey Optimization (SMO), which leverages the fission-fusion social structures of spider monkeys to balance exploration and exploitation while effectively navigating local and global optimal solutions. Avinash Sharma et al. have proposed an Ageist SMO (ASMO) based on a spider monkey population's age difference [33], [34].

Wang et al. [35] have proposed the Elephant Herding Optimization (EHO) algorithm, inspired by elephant herding behavior, featuring a clan updating operator and a clan separating operator for updating the positions of elephants and matriarchs within clans.

The Virus Optimization Algorithm (VOA) simulates virus attacks [36]. It will be further improved in 2020 by reducing the number of parameters that need to be set and hence making it self-adaptive [37].

Salehan and Deldari created Coronavirus optimization (CVO), miming COVID-19 behaviour [38]. SIR mathematical models are used to analyze COVID-19 epidemic behaviour. CVO is a feasible and effective approach for resolving application issues. The limitation of CVO includes a lack of implementation to optimize real-time problems and several different control parameters. In [39], the PID controller was designed for brushless DC (BLDC) motors with several disturbances by optimizing CVOA.

Mirjalili et al.[40] have developed the Salp Swarm Optimization (SSO) method, which organizes a randomly generated population into leaders and followers to optimize through exploration and exploitation phases, inspired by deep-sea salp behavior.

Arora and Singh [41] have developed the butterfly optimization algorithm (BOA), based on butterflies' natural behavior in foraging and mating. Despite global and local search capabilities, BOA may experience slow convergence.

Shadravan et al. [42] have developed the Sailfish Optimizer (SFO), inspired by hunting sailfish, using sailfish for intensification and sardines for search space diversification. SFO balances these strategies to prevent early convergence.

Hayyolalam and Kazem [43] have proposed the black widow optimization (BWO) algorithm in 2020, inspired by black widow spider mating behaviour. BWO offers good results, is fast-converging, and avoids local optima.

The Sparrow search algorithm (SSA), developed by Xue and Shen in 2020 [44], optimizes sparrows' foraging, predatory, and anti-predatory behavior, but struggles with insufficient stimulation and stagnant exploitation due to poor trade-offs.

Hubálovská et al.[45] have proposed the Botox Optimization Algorithm (BXOA), a novel human-based metaheuristic algorithm that simulates the effects of Botox injections on the human body. The limitations of BXOA include limited applicability, parameter tuning is complex, and there is a risk of premature convergence.

Al-Baik et al. [46] developed the Pufferfish Optimization Algorithm (POA), a bio-inspired metaheuristic mimicking pufferfish behavior, with exploration and exploitation phases for efficient high-dimensional optimization. However, POA has limitations, such as no guarantee of finding a global optimum and no assumptions about implementation success.

In [47], the authors proposed the Coronavirus Optimization Algorithm (CVOA) to model virus spread and infection, avoiding arbitrary initialization and iteration termination. However, it faces challenges like exponential growth of infected populations and the absence of a candidate reduction mechanism, affecting performance and convergence.

The Novel COVID-19 Based Optimization Algorithm (C-19BOA) [48], uses bio-inspired methods to improve power system performance. Inspired by the virus's spread and adaptation, it incorporates social distancing, mask use, and antibody rate, revolutionizing problem-solving by leveraging nature's strategies.

Khalid et al. [49] have proposed MOCOVIDOA, a novel algorithm for multi-objective optimization inspired by the dynamics of the coronavirus, enabling simultaneous resolution of multiple objectives. This promising approach efficiently addresses complex optimization challenges across various domains, using the Pareto optimal dominance operator for solution assessment.

Saxena et al. [50] introduced the Marine Predator Chaotic Algorithm (MPCA), a novel method based on the Marine Predator Algorithm (MPA) and the $\beta$ chaotic map. The position update method, utilizing a fusion of chaotic functions, improves MPA performance and is evaluated using a COVID-19 dataset.

Selim et al. [51] have proposed the Modified Coronavirus Herd Immunity Aquila Optimization (MCHIAO), a hybrid optimization technique that improves feature selection tasks by classifying cases, refining gene values using coronavirus-inspired chaotic systems, and alternating between expanded and narrowed exploitation.

A Coronavirus Mask Protection Algorithm is proposed in [52],simulates human protective measures against the virus, consisting of three stages: infection, diffusion, and immune phases. A parallel version is suggested to handle multiple virus strains, achieving better results in fewer iterations.

In 2024, Chu et al. [53] proposed the Ship Rescue Optimization (SRO) algorithm, inspired by ship maneuvering and rescue dynamics. SRO models have two search strategies—delayed (large-scale) and immediate (focused) to efficiently solve optimization problems using ship motion equations for position updates.

In 2024, Dou et al. [54] proposed the Binary Runge Kutta Optimizer (GBRUN) with Gradient Search to address high-dimensional feature selection problems. By incorporating S-, V-, and U-shaped transfer functions, GBRUN transforms the continuous RUN into a binary version, enhancing exploration and outperforming other advanced algorithms in feature selection and classification accuracy.

## III. Social Distancing Induced Coronavirus Optimization Algorithm (COVO)

### A. *Inspiration*

One of the most critical challenges of the 21st century is to prevent SARS-CoV-2 and COVID-19 diseases [55]. The virus migrated to multiple nations, forcing the World Health Organization (WHO) to designate it as a new contagious disease. Coronavirus cases are increasing rapidly as shown in Figure 1. Figure 1 shows the total number of COVID-19 cases surpassing 704 million by May 2024, as reported by WHO [56]. The data indicates steady growth, with significant increases until mid-2022, followed by a plateau. Though vaccinations and medications are available for COVID-19, identifying the infected patients earlier is essential to treat them immediately and prevent the spreading of the virus to others. COVID-19 has a mortality rate that varies from country to country between 0.25 and 3.0 per cent [57]. Figure 2 shows a sharp rise in COVID-19 deaths from late 2020 to 2021, with a slower increase from early 2022. By May 2024, total deaths approach 7.5 million. Natural immunity refers to a large number of people in a society who are immune to illness (either by vaccination or spontaneous infection), and as a result, the pathogen is unable to disseminate. It occurred since more than 60% of the community had survived the illness (herd immunity threshold). Natural immunity can potentially impact pandemic dissemination by slowing the spread of disease. One of the strategies recommended to prevent the COVID-19 pandemic breakout is herd immunity. It's important to note that this method employs Darwin's notion of the survival of the fittest. According to the social distancing theory, COVID-19 cannot be transmitted from person to person if an individual has not been in close contact with an infected person (within 1.8 meters). As a result of social distancing, direct contact between people is avoided, and the probability of spreading virus-carrying droplets is reduced [58]. The government used two strategies to prevent the spread of COVID-19, lockdown in the country and social distancing. Susceptible refers to a person who isn't immune to the virus. When a person is infected with COVID-19, he can be disseminating a case. Consequently, depending on the strength of the person's immune system, he could recover (i.e., immune). Generally, the immune system of aged people is weaker than those of young people as they are more likely to have various illnesses such as diabetes, cardiovascular disease, or cancer. Hence, a person's age has a significant impact on whether or not they can recover from this disease. The key steps for developing the COVO algorithm are as follows.

• A massive group of sick individuals infects another significant set of individuals.

• The majority of those who were infected have healed, and only a few have died.

• After some time, the majority of people will be immune to the disease.

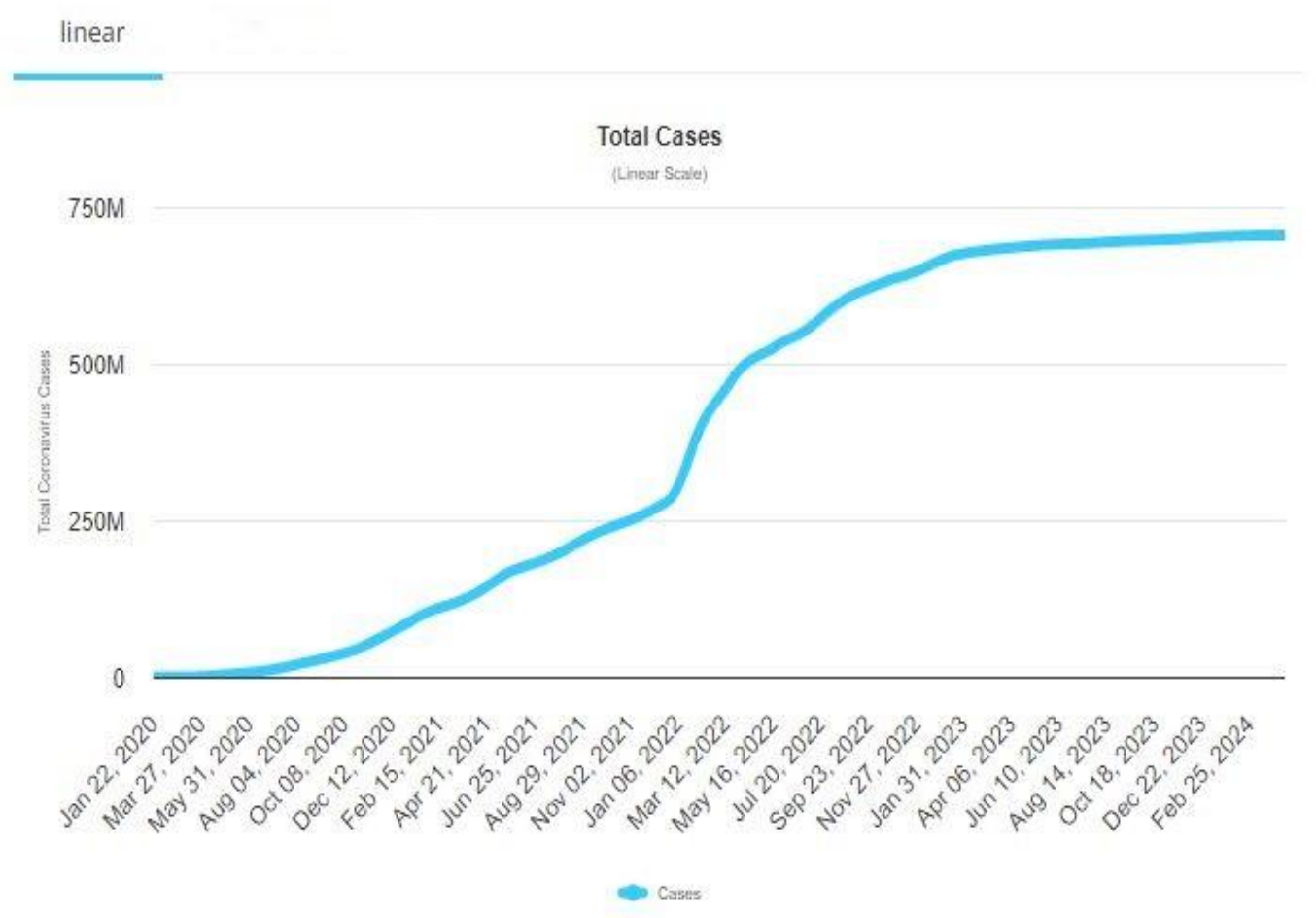

Fig. 1 Total Coronavirus Cases from January 2020 to May 2024 [59]

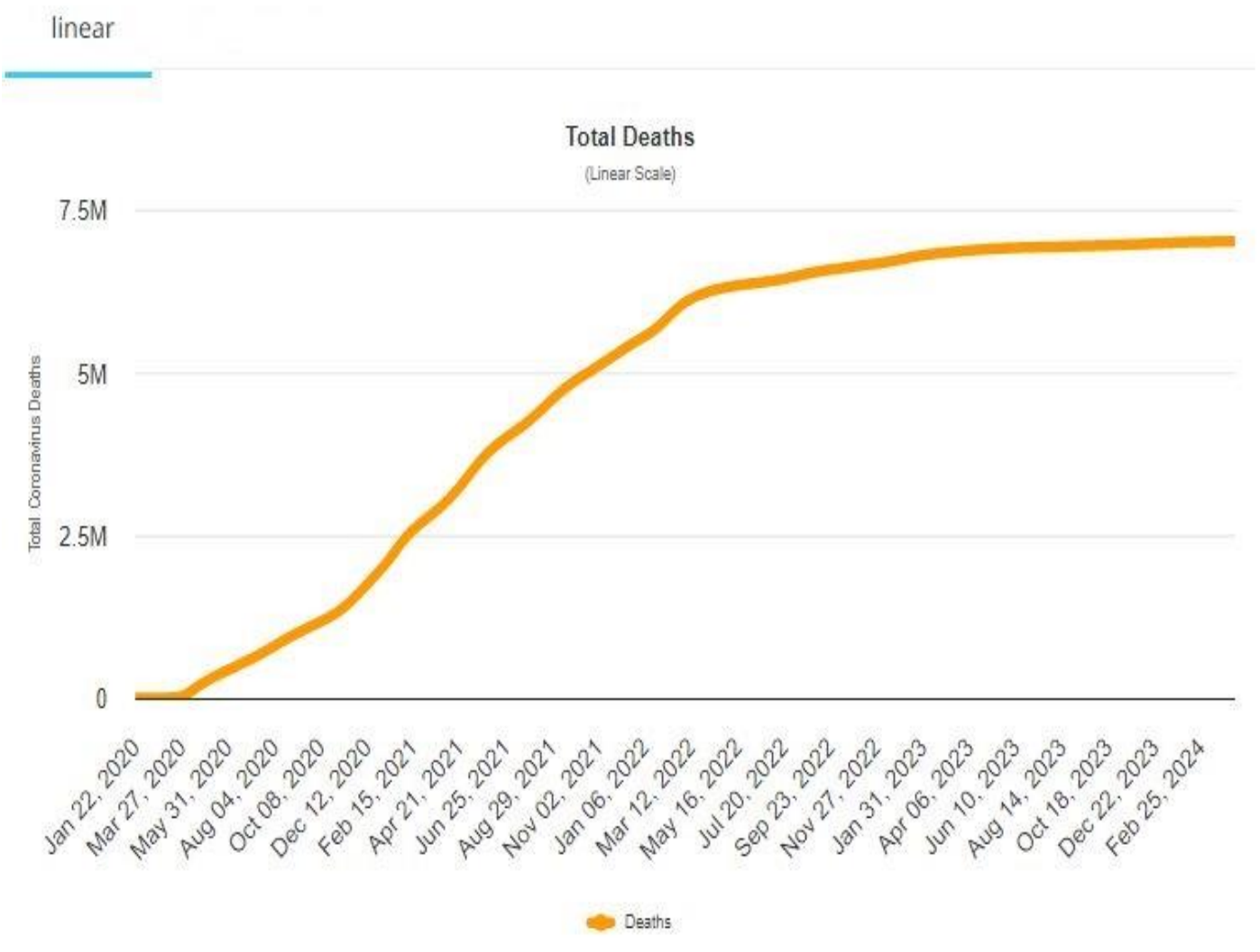

Fig. 2 Total Coronavirus Deaths from January 2020 to May 2024 [59]

The SARS-CoV-2 virus causes COVID-19, which may affect persons of all ages, particularly those with a history of medical problems. COVID-19 has a reproductive number (R0) of around 2.5, as per a WHO study [60]. It shows that at least every sick individual infects 1 to 3 individuals. COVID-19 is disseminated in two ways: by inhaling and through close association. COVID-19 cannot be transmitted by maintaining a physical distance between two persons of at least 1.8 meters (6ft).
Figure 3 shows the impact of social distancing on COVID-19 infection in scenario 1, with social exposure reduced by 50%, and in scenario 2, where social exposure was reduced by 75%.

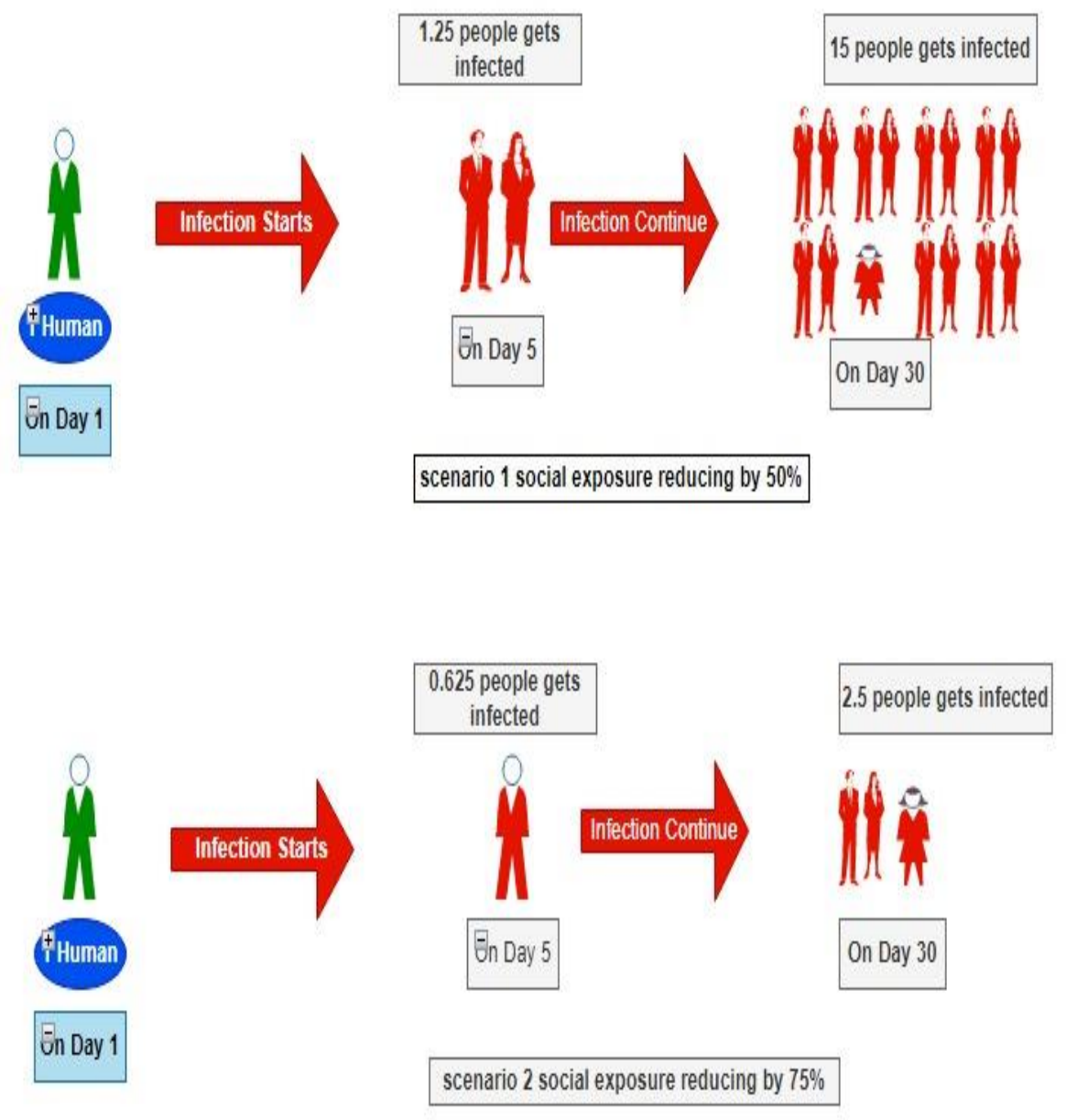

Fig. 3 Impact of Social Distancing on COVID-19 Infection; In Scenario 1, Social Exposure Was Reduced by 50%, and In Scenario 2, Social Exposure Was Reduced by 75% [1].

Alternative techniques, like face masks, may reduce viral transmission and the outbreak. According to previous studies, social distance is among the most effective methods for reducing COVID-19 transmission. Social distancing significantly impacts COVID-19 transmission, with a 50% reduction leading to meaningful decreases in infection rates and easing healthcare burdens. A more challenging 75% reduction can further lower infection rates and help control outbreaks more quickly, underscoring the importance of social distancing in managing pandemics. The goal of this technique appears to be to keep everyone in society healthy while also rehabilitating and promoting the health and well-being of those infected. After the epidemic, it is indeed expected that everyone will be healthy.

### B. *Methodology*

A novel Social Distancing Induced Coronavirus Optimization (COVO) algorithm is proposed in this research work. It is introduced to reduce the spreading rate of the virus around the globe. The social distancing parameter is integrated with the COVO model to reduce or interrupt the transmission of COVID-19 amongst the population.

Algorithm 1 shows the pseudo-code of the COVO algorithm. COVO algorithm replicates WHO-recommended methods for reducing COVID-19 disease transmission and improving public health. The WHO and various countries have implemented a few mitigating strategies in response to COVID-19. These practices include quarantine, isolation, personal cleanliness, wearing plastic screens and face masks in public, restricting people's movement, and avoiding large-group public meetings. The social distancing strategy's goal is to maintain a safe gap between people, reducing the chances of getting an infection. The COVO attempts to tackle optimization constraints by simulating social distance, a one-stop mitigation technique. COVO's demonstrated strengths in both multimodal function optimization and noise removal suggest that it can be a valuable tool for a range of optimization problems. Its performance improvements over other algorithms make it a preferable choice for applications requiring reliable and efficient optimization solutions. The goal is to find the population's fittest and healthiest individual, which correlates to a near-optimal solution for an optimization problem.

Step 1: Initialization of population and parameters

The population $pop$ is initialized as follows.

$$pop = [p_1, p_2 ...., p_N]$$

Here, each solution $p_i \in pop$ in the population is a person and $N$ represents the size of the population. In addition, the parameters are initialized as specified in TABLE I. $P_{die}$ and $D_{rate}$ are generated using the random chaotic map [61] using Eq. (1) and Eq. (2), respectively.

$$P_{die} = abs(P_{die} + b - abs(p - 2\pi)) * np.\sin(2\pi * P_{die}) \quad (1)$$

$$D_{rate} = abs(D_{rate} + b - abs(p - 2\pi)) * np.\sin(2\pi * D_{rate}) \quad (2)$$

Within this $pop$, only one individual is referred to as a zero-infected patient (PZ). This individual identifies the first infected person. If there are no identified previous local minima, then PZ is suggested to undergo random initialization.

TABLE I COVO PARAMETERS [47]

| Symbol | Description | Initial value |
|---|---|---|
| $S_{rate}$ | Spreading Rate | a random value between 0 to 0.5 |
| $SS_{rate}$ | Super Spreading Rate | a random value between 0.5 to 1 |
| $P_{travel}$ | Probability of travel | random binary value 0 or 1 |
| $P_{die}$ | Probability of Death | 0.43597 |
| $D_{rate}$ | Death Rate | 0.13955 |
| $H_{dist}$ | Social distancing parameter | 13.77323 |
| T | Threshold value | 0.5 |
| L | Lower bound value | -10 |
| U | Upper bound value | 10 |
| $\Delta$ | Constant parameter | 0.97248 |
| $dist_{ij}^t$ | Distance between two persons *i* and *j* at time instant *t* | 13.77323 |
| $Fit$ | Fitness Function | 21.43888 |

Step 2: The search vectors $X_{ij}$ are initialized randomly. Here $i = 1,2,..., m; j = 1,2,..n$. Here, $m$ denotes the number of search agents and $n$ indicates the dimension length.
Step 3: The opposite solutions are generated using the Opposition learning behaviour [62].
Step 4: Disease propagation
The valuation of the diverse cases takes place based on the individual.

(a) Based on the death rate for COVID-19, each infected individual is said to have a dying probability $P_{die}$. With this $P_{die}$, the infected population expires, and the case fatality ratio varies in terms of age, health condition, and locality. From these individuals, COVID-19 couldn't spread to others.

(b) The live infected persons spread the virus to new individuals (intensification). As per the given probability $P_s$ (spreader), two sorts of spreading categories are considered: ordinary spreaders and super-spreaders.

(i) Ordinary spreaders: New individuals are affected by the infected population based on the regular spreading rate $S_{rate}$.

(ii) Super-spreaders: Based on the super-spreading rate $SS_{rate}$, the new individuals are affected by the infected individuals. Once the infected individual becomes a super-spreader, he can share the virus with 15 healthy persons.

(c) To ensure diversification, the super-spreader and ordinary individuals could travel in the search space to explore new solutions. The travelling probability $P_{travel}$ is available among these infected individuals, and using this $P_{travel}$, they propagate the disease and spread it to travellers with different travelling rates $T_{rate}$.

✓ For zero-infected patients (PZ), If $P_{travel}$ =0, the solutions are updated by verifying two criteria.

(i) If the social distancing parameter $H_{dist}$ is lower than the threshold value $T$ (i.e., $H_{dist} < T$), then update the solution using Eq. (3).

$$X = L + (U - L) \tag{3}$$

Here $L$ and $U$ are the upper and lower bounds of the solutions.

(ii) If the social distancing parameter $H_{dist}$ is greater than the threshold value $T$ (i.e., $H_{dist} > T$), then update the solution using Eq. (4).

$$X = L + (U - L).S_{rate} \tag{4}$$

Here, $H_{dist}$ the social distancing parameter shows the minimum safe distance between two individuals $X_i$ and $X_j$. This operator simulates the social distancing policy. Obviously, not all people engage in social distancing at any given time, but only a proportion of the population follows this policy. As the disease spreads, the importance of social distancing becomes more apparent, and people are forced to be more observant.

$$H_{dist_{ij}} = \begin{cases} \Delta - dist_{ij}^t & if\ dist_{ij}^t < \Delta \\ dist_{ij}^t & if\ dist_{ij}^t \geq \Delta \end{cases} \tag{5}$$

Here, the current distance between two persons at instance t is computed as $dist_{ij}^t = \left| X_i^t - X_j^t \right|$

✓ For zero infected patients (PZ), if $P_{travel}$ =1, then update the solution using Eq. (6).

$$X = L + (U - L).SS_{rate} \tag{6}$$

Step 5: Fitness Computation

Compute the fitness $Fit$ of all the individuals using Eq. (7). The fitness function $Fit$ is fixed as a minimization of error $E$ and this $E$ includes the 13 benchmark functions $f_1 - f_{13}$.

$$Fit = \min(E) \tag{7}$$

Step 6: Population updating Based on this $P_{die}$, the solution is updated is carried out.

- If $Fit > P_{die}$, then a newly infected patient is identified and updated using Eq. (8). Then, move to Step 7.

$$X_{new} = X_{old} \pm X_{old}[Fit.P_{die}] \tag{8}$$

- If $Fit \leq P_{die}$ this patient is said to be dead, newly infected patients are generated by proceeding to Step 2.

Three different populations (Deaths, Recovered population, and Newly infected population) are considered for each generation.

(a) Deaths- If an infected person expires, they are added to this population but are not considered again.

(b) Recovered population- The infected individuals who recover are transmitted to the recovered population. These recovered individuals may get re-infected with a probability $P_{reinfected}$. The people within this population might get affected over any iteration, provided that the re-infection criterion is satisfied. On the other hand, the isolated individuals maintaining social distance remain in recovered population if the isolation probability $P_{isolation}$ is satisfied. $P_{isolation}$ is uncertain as it varies from country to country. After each iteration, the count of the infected population rapidly lessens by following various social isolation measures.

(c) New infected population- At every iteration, all the individuals affected by the virus are considered. The number of repeated new individuals increases with each iteration, so it is recommended to eliminate such repeated individuals before proceeding to the next iteration.

Step 7: If $X_{new} = X_{old}$; then the newly infected population cannot infect the new individuals. So, move to the termination stage. Otherwise repeat step 2.

Termination- One of the most intriguing aspects of the suggested method is its capacity to terminate without requiring any parameters. This scenario arises because the healed and diseased populations continue to expand over time, while the newly afflicted community has been unable to infect new people. The number of infected persons is predicted to rise for a given amount of repetitions. Despite this, since the healthcare infrastructure is too large, and the diseased population deteriorates with time, the number of infected people will indeed be lower than the current total population at a certain point in time. In addition, the stop criterion can be adjusted to provide a specified number of iterations. The social distance idea assists in meeting the stopping requirement as well. Social distancing is used to update the population until terminating requirements are satisfied. Ultimately, the person with the best fitness will be introduced as the best solution to the situation. Figure 4 shows the flowchart for the Social Distancing Induced Coronavirus Optimization (COVO) Algorithm.

**Algorithm 1: Social Distancing Induced Coronavirus Optimization (COVO) Algorithm**

1. Initialize the size of population, population, probability of death, death rate, spreading rate, super spreading rate, social distance parameter, probability of travel and maximum iterations.
2. Initialize the search agents randomly, $j = 1,2,...m; i = 1,2...n.$
3. Opposite solutions are generated using the opposition learning behavior.
4. If $P_{travel}$ =0
5.     If $H_{dist} < T$
6.         Zero infected patient is computed as
   $$X = L + (U - L)$$
7.     Else
8.         Zero infected patient is calculated as
   $$X = L + (U - L).S_{rate}$$
9.     End if
10. Else
11.         Zero infected patient is computed as
   $$X = L + (U - L).SS_{rate}$$
12 End if
13. Fitness is computed for all the individuals
   $$Fit = \min(E)$$
14. If $Fit > P_{die}$
15.     New infected patient will be
   $$X_{new} = X_{old} \pm X_{old}[Fit.P_{die}]$$
16. Else
17.     The patient is dead and generate a new infected patient.
   Proceed to step 2.
18. End if
19. If $((Xnew == Xold) \,||\, (iter < \max iter))$
20.     The newly infected population cannot infect the new individual. Move to termination stage.
21. Else
   Go to step 2
22. End if

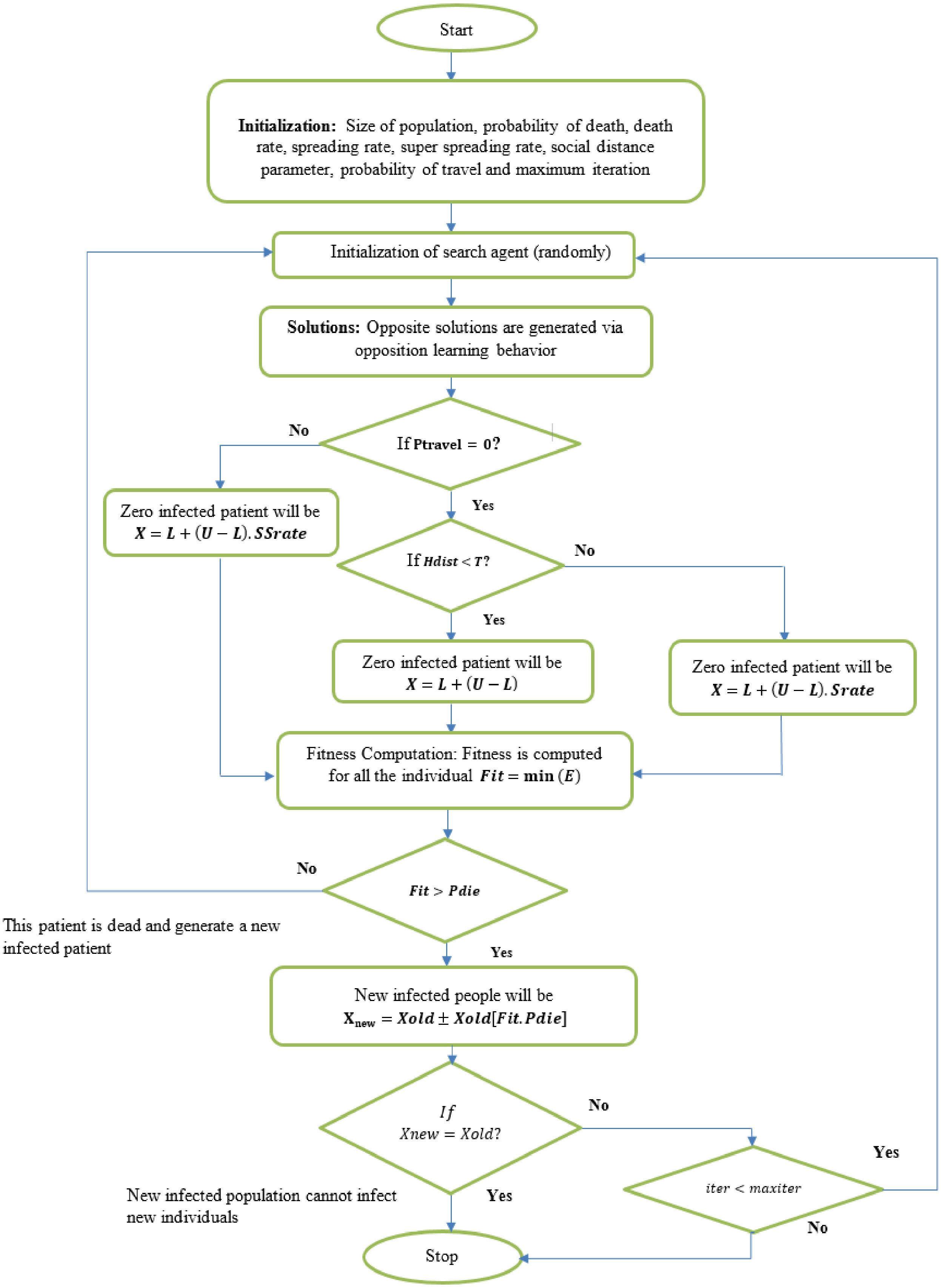

Fig. 4 Flowchart of Social Distancing Induced Coronavirus Optimization Algorithm (COVO).

TABLE II COVO INITIAL PARAMETERS VALUES [47]

| Symbol | Description | COVO initial values |
|---|---|---|
| N | Population Size | 10 |
| $S_{rate}$ | Spreading Rate | 0.3575 |
| $SS_{rate}$ | Super Spreading Rate | 0.80138 |
| $P_{travel}$ | Probability of travel | 0 |
| $P_{die}$ | Probability of Death | 1 |
| $D_{rate}$ | Death Rate | 0.888557 |
| $H_{dist}$ | Social distancing parameter | 0.87306 |
| T | Threshold value | 0.5 |
| L | Lower bound value | -100 |
| U | Upper bound value | 100 |
| $\Delta$ | Constant parameter | 0.41466 |
| $dist_{ij}^{t}$ | Distance between two persons at a time instant 't' | 0.87306 |
| $Fit$ | Fitness Function | 0.19370 |

## IV. RESULTS AND DISCUSSIONS

The suggested COVO algorithm is implemented in Python, and 13 typical benchmark functions $(F_1 - F_{13})$ have been used during the experimental investigation to evaluate the algorithm's performance. These benchmark functions $(F_1 - F_{13})$ and their corresponding range are shown in TABLE III. Here $(F_1 - F_5)$ are functions having unimodal properties. The function $F_6$ is a step function with one minimum and a discontinuous function. $F_7$ is a quartic noise function with uniformly distributed random variables between 0 to 1. $(F_8 - F_{13})$ are multimodal functions, wherein the local minima expand exponentially concerning the dimension of the problem. In most of the functions $(F_1 - F_4)$ (unimodal) $(F_6 - F_7)$ and $(F_9 - F_{11})$ (multimodal), the COVO Algorithm has achieved higher convergence. The proposed model's effectiveness is evaluated by comparing it with several existing models using the best, median, worst, standard deviation, and mean square error.

TABLE III STANDARD NUMERICAL BENCHMARK FUNCTIONS AND THEIR RANGE

| Sr. No. | Function (F) | Range of the function |
|---|---|---|
| 1) | $F_1 = \sum_{i=1}^{d} x_i^2$ | $[-100,100]^D$ |
| 2) | $F_2 = \sum_{i=1}^{d} \lvert x_i \rvert + \prod_{i=1}^{d} \lvert x_i \rvert$ | $[-10,10]^D$ |
| 3) | $F_3 = \sum_{i=1}^{d} \left( \sum_{j=1}^{i} x_i \right)^2$ | $[-100,100]^D$ |
| 4) | $F_4 = \max_i \{ \lvert x_i \rvert, 1 \le i \le d \}$ | $[-100,100]^D$ |
| 5) | $F_5 = \sum_{i=1}^{d-1} [100(x_{i+1} - x_i^2)^2 + (x_i - 1)^2]$ | $[-30,30]^D$ |
| 6) | $F_6 = \sum_{i=1}^{d-1} [\lvert x_i + 0.5 \rvert]^2$ | $[-100,100]^D$ |
| 7) | $F_7 = \sum_{i=1}^{d} i x_i^4 + rand[0,1]$ | $[-1.28,1.28]^D$ |
| 8) | $F_8 = 418.9829 * D - \sum_{i=1}^{d} (x_i \sin(\sqrt{\lvert x_i \rvert}))$ | $[-500,500]^D$ |
| 9) | $F_9 = \sum_{i=1}^{d} x_i^2 - 10\cos(2\pi x_i) + 10$ | $[-5.12,5.12]^D$ |
| 10) | $F_{10} = -20\exp\left(-0.2\sqrt{\frac{1}{d}\sum_{i-1}^{d} x_i^2}\right) - \exp\left(\frac{1}{d}\sum_{i-1}^{d}\cos(2\pi x_i)\right) + 20 + \exp(1)$ | $[-32,32]^D$ |
| 11) | $F_{11} = \frac{1}{4000}\sum_{i=1}^{d} x_i^2 - \prod_{i=1}^{d}\cos\left(\frac{x_i}{\sqrt{i}}\right) + 1$ | $[-600,600]^D$ |
| 12) | $F_{12} = \frac{\pi}{d}\left\{10\sin^2(\pi y_1) + \sum_{i=1}^{d}(y_i - 1)^2[1 + 10\sin^2(\pi y_{i+1})] + (y_n - 1)^2\right\} + \sum_{i=1}^{d} u(x_i,10,100,4);$ <br> $where\ y_i = 1 + \frac{1}{4}(x_i + 1);$ <br> $u(x_i,a,k,m) = \begin{cases} k(x_i - a)^m & x_i > a \\ 0 & -a \le x_i \le a \\ k(-x_i - a)^m & x_i < -a \end{cases}$ | $[-50,50]^D$ |
| 13) | $F_{13} = 0.1\left\{\sin^2(\pi 3x_1) + \sum_{i=1}^{d}(x_i - 1)^2[1 + \sin^2(3\pi x_{i+1})] + (x_n - 1)^2[1 + \sin^2(2\pi x_d)]\right\} + \sum_{i=1}^{d} u(x_i,5,100,4);$ <br> $u(x_i,a,k,m) = \begin{cases} k(x_i - a)^m & x_i > a \\ 0 & -a \le x_i \le a \\ k(-x_i - a)^m & x_i < -a \end{cases}$ | $[-50,50]^D$ |

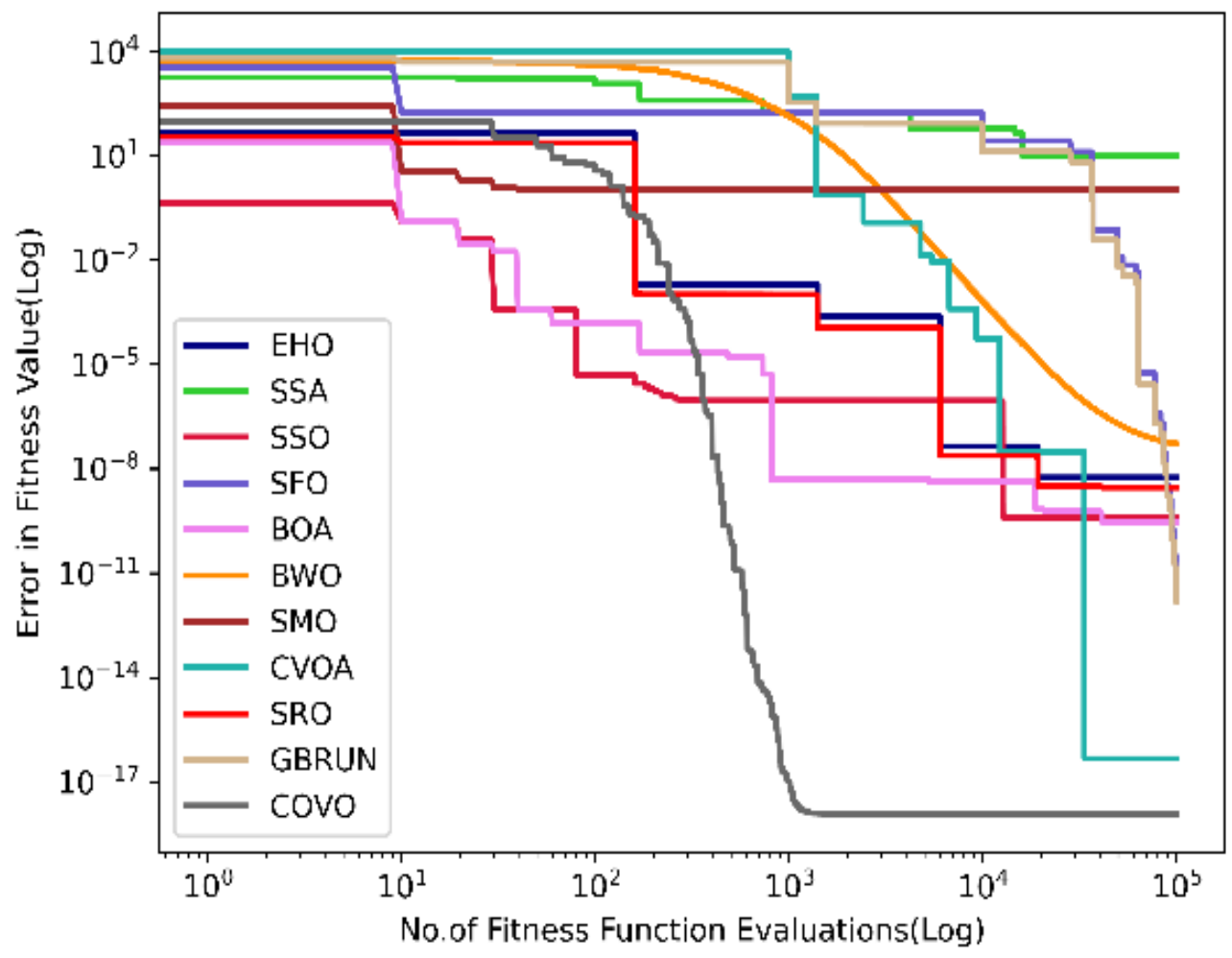

(a)

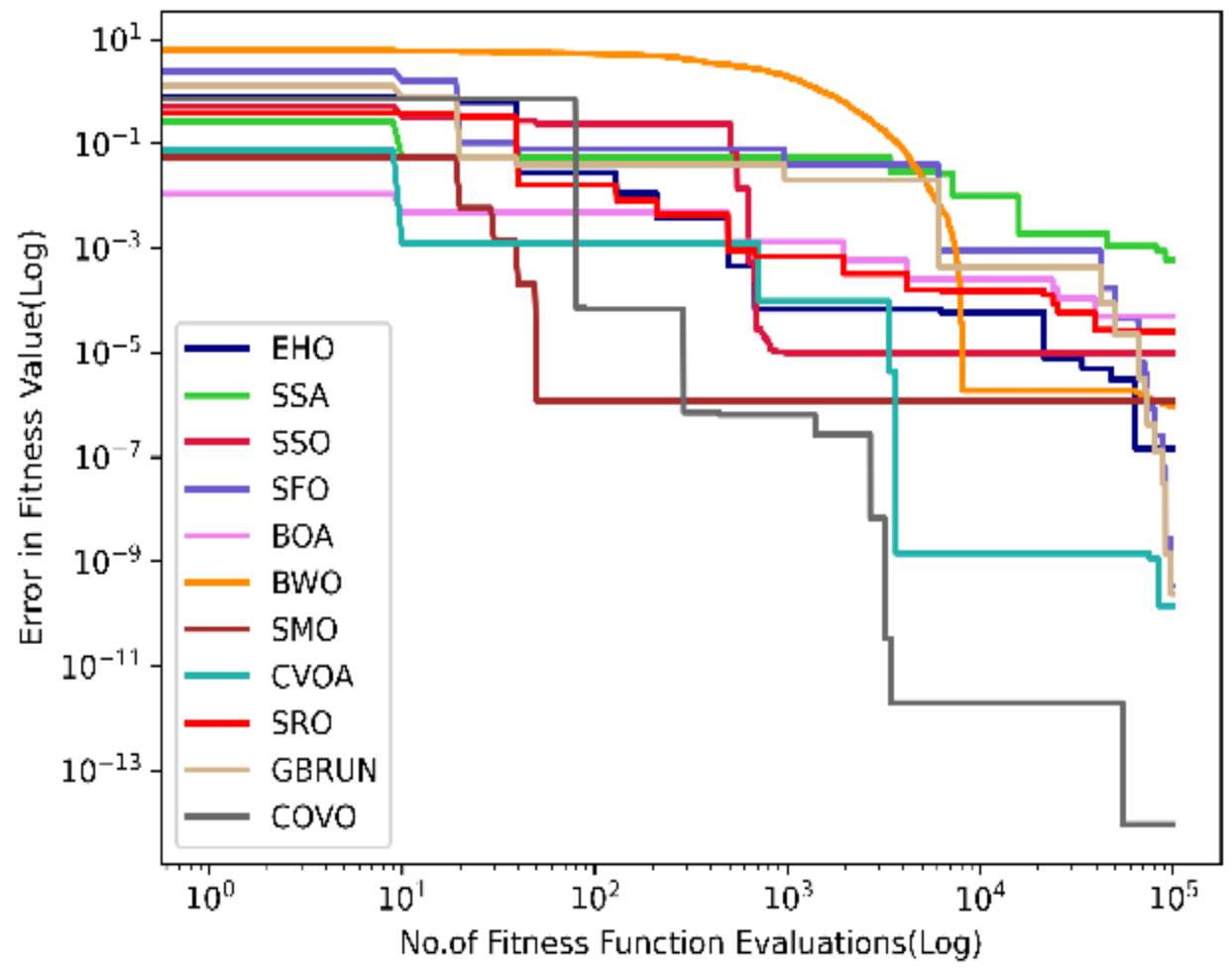

(b)

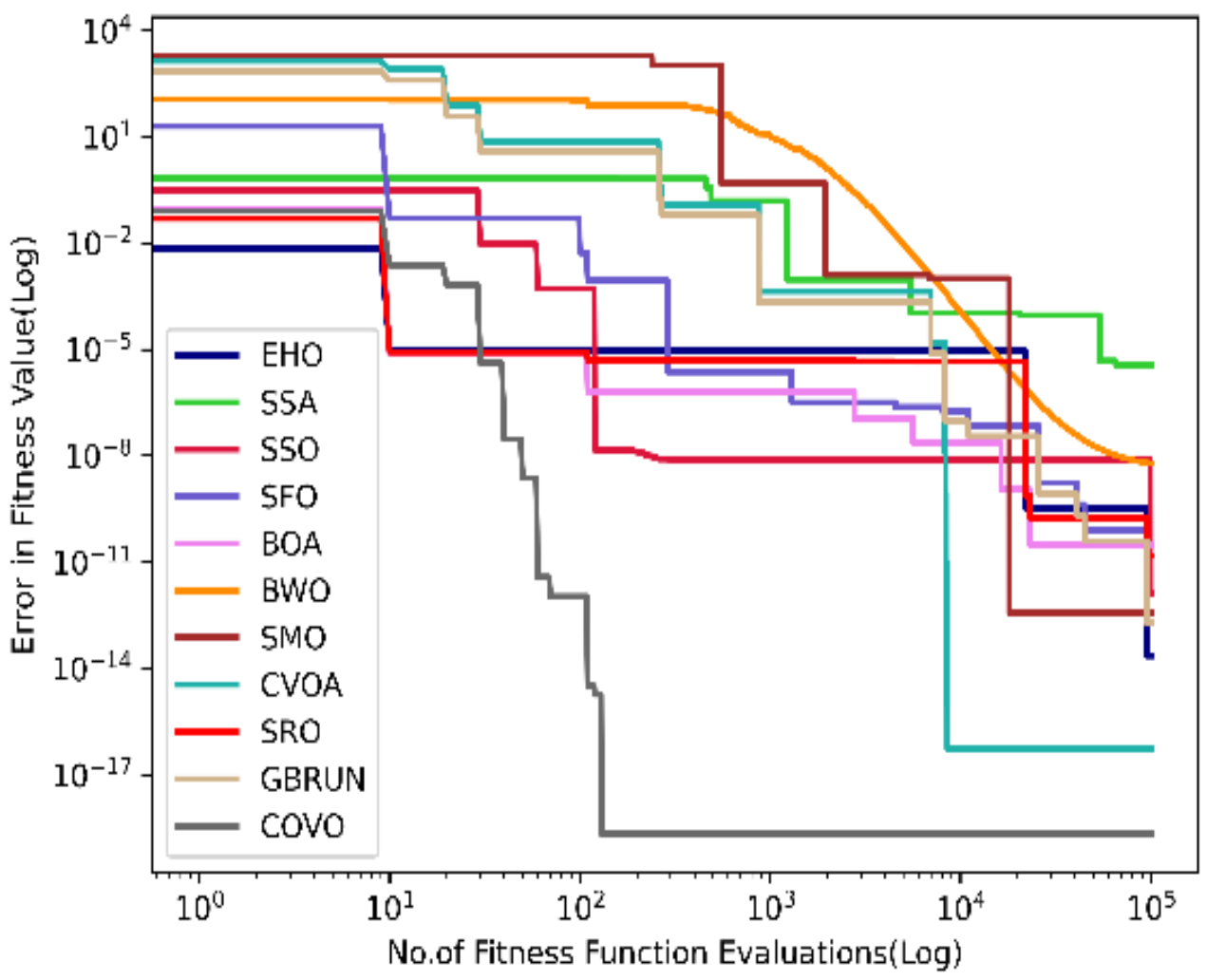

(c)

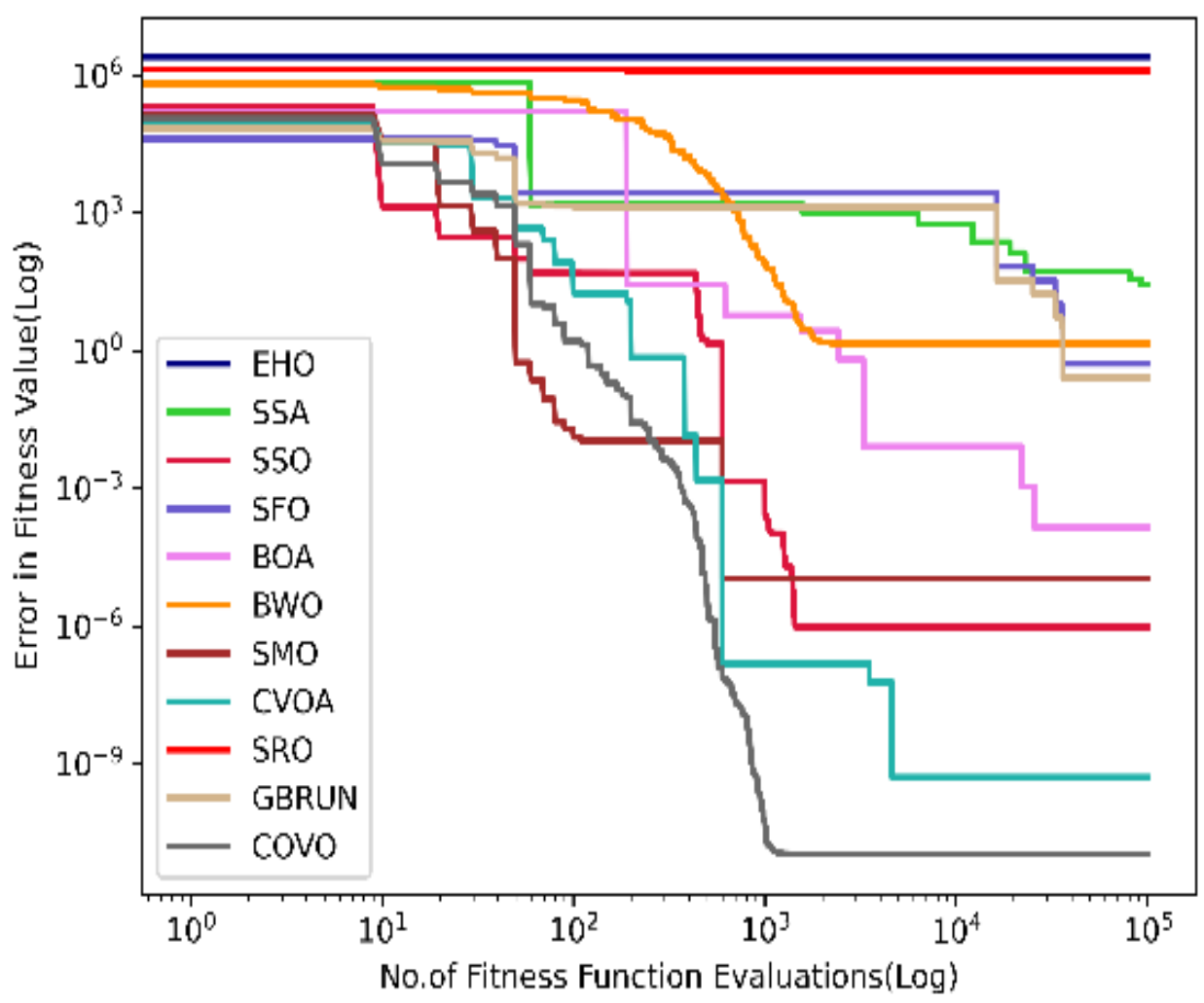

(d)

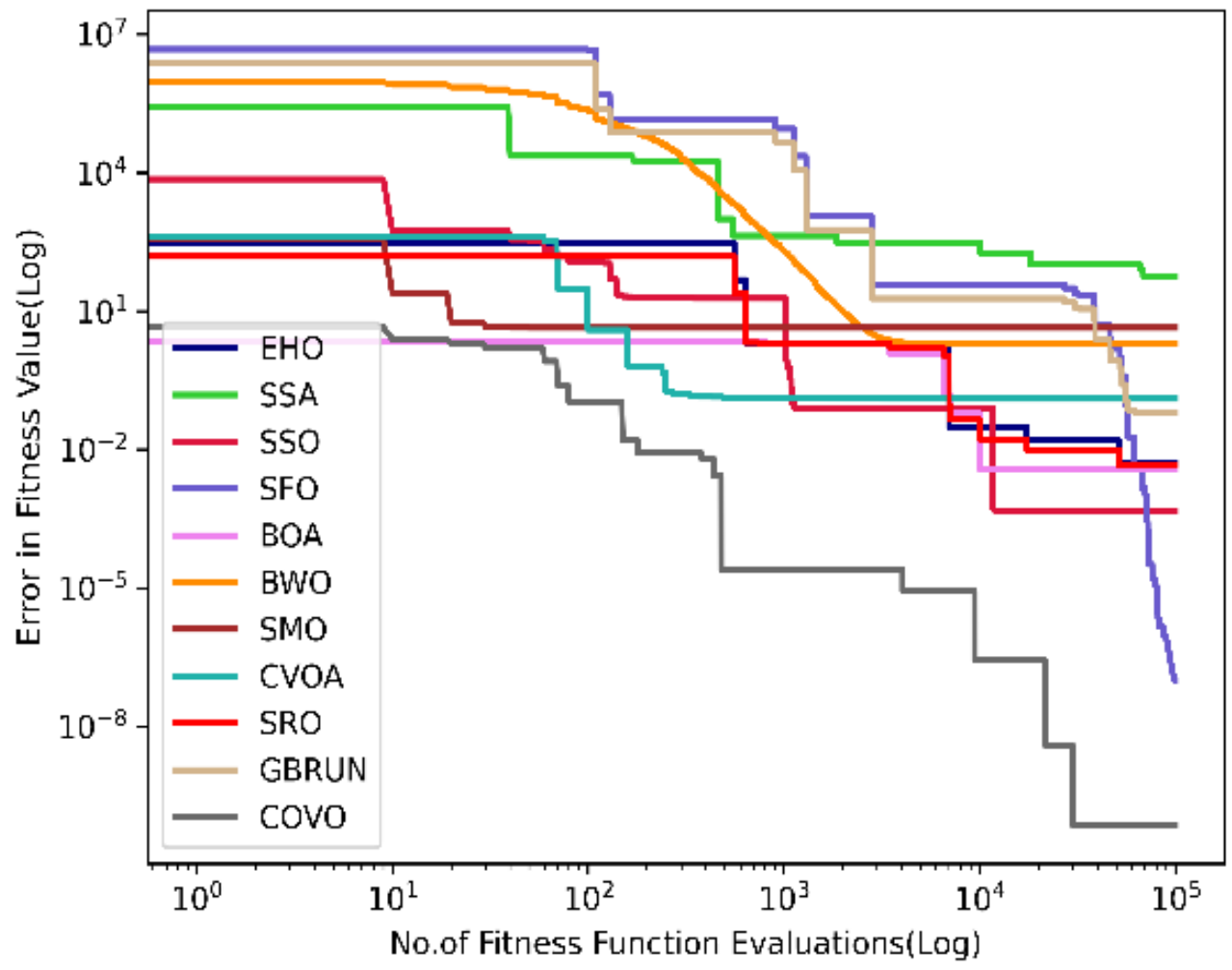

(e)

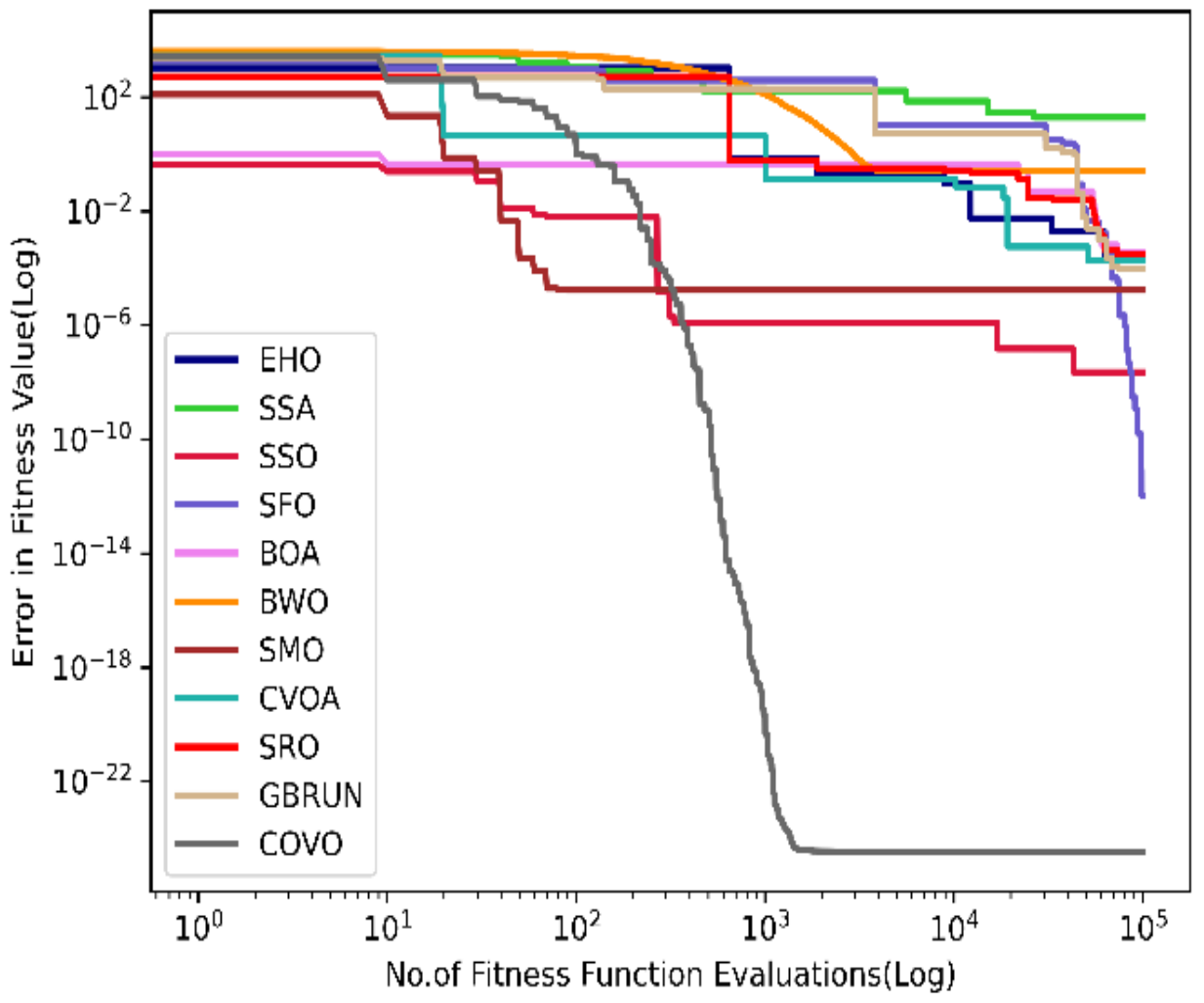

(f)

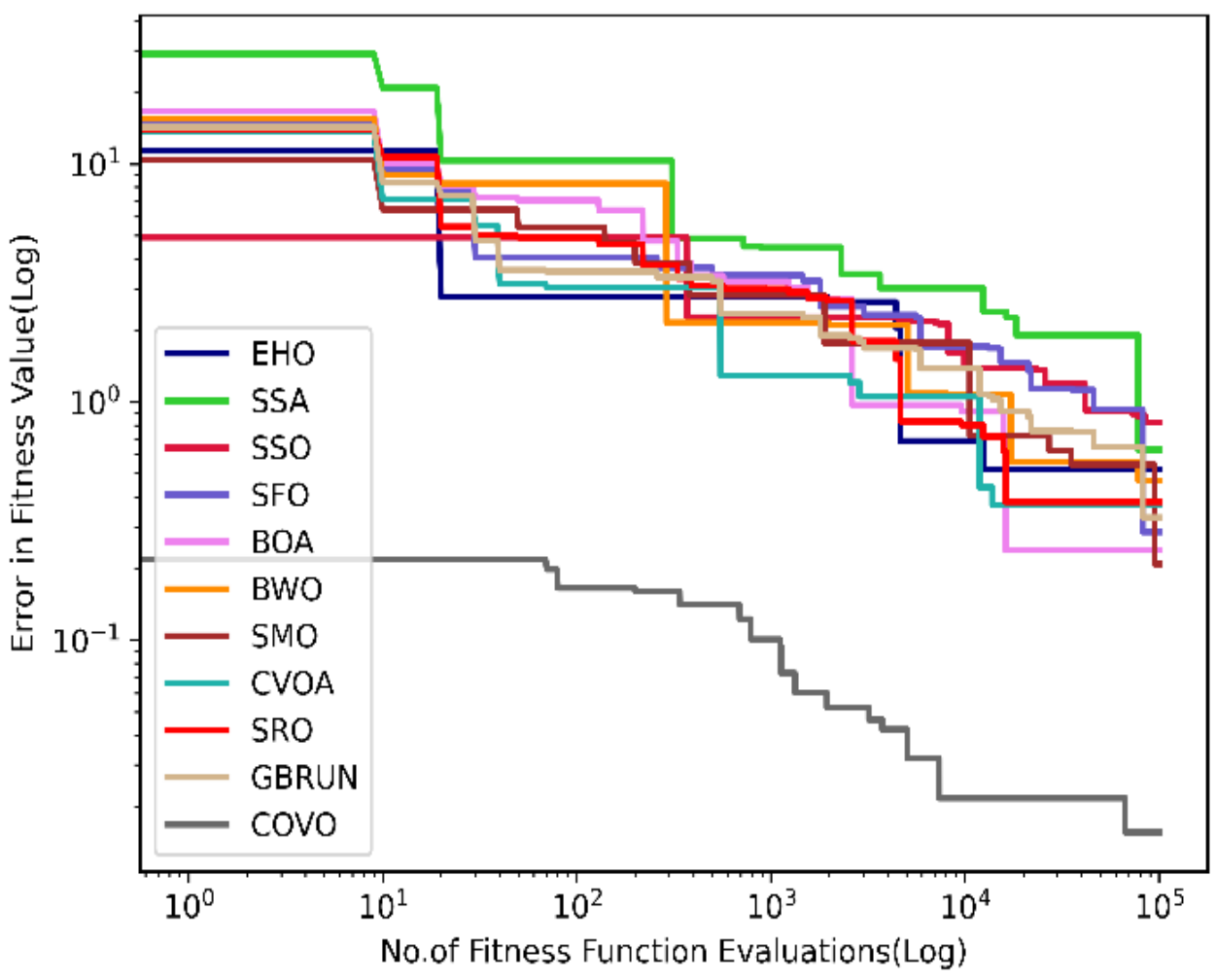

(g)

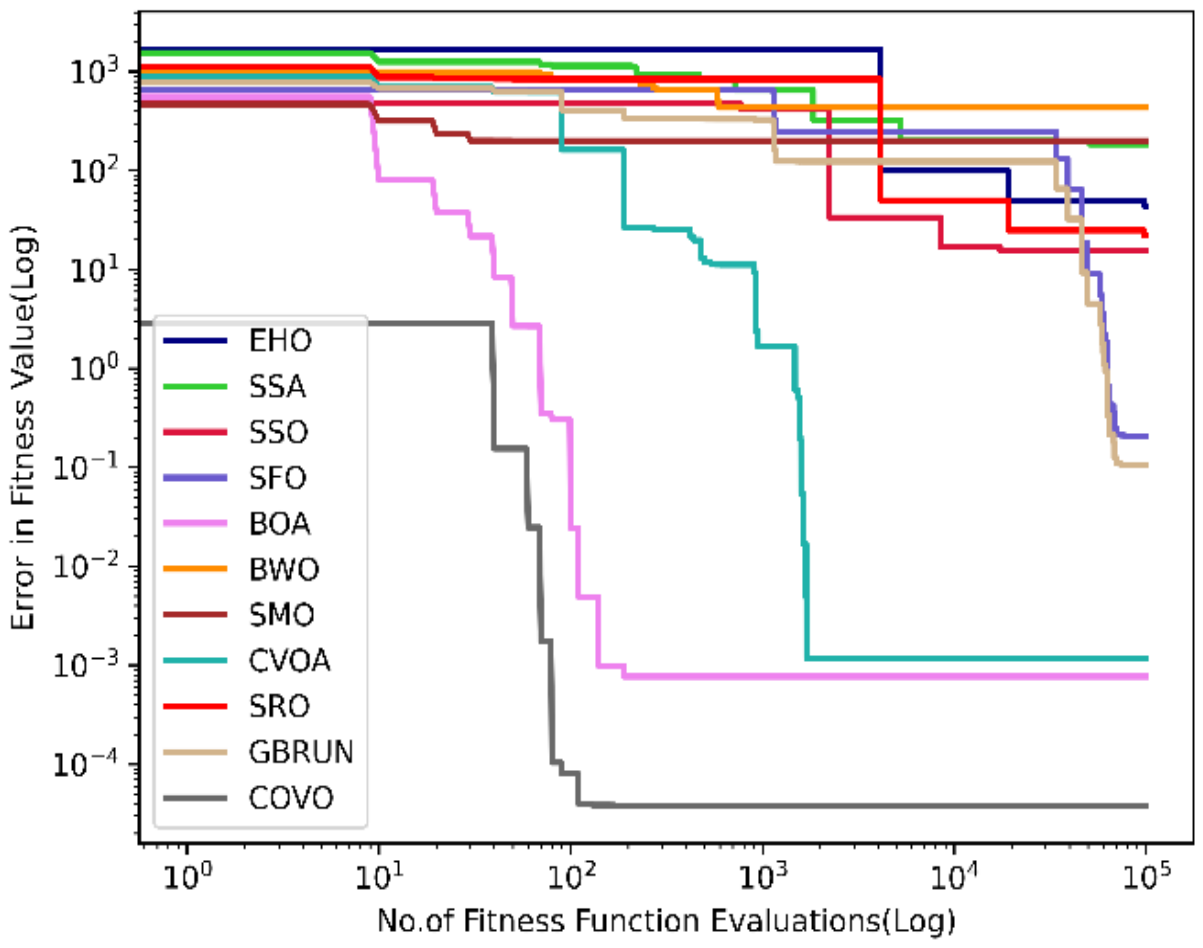

(h)

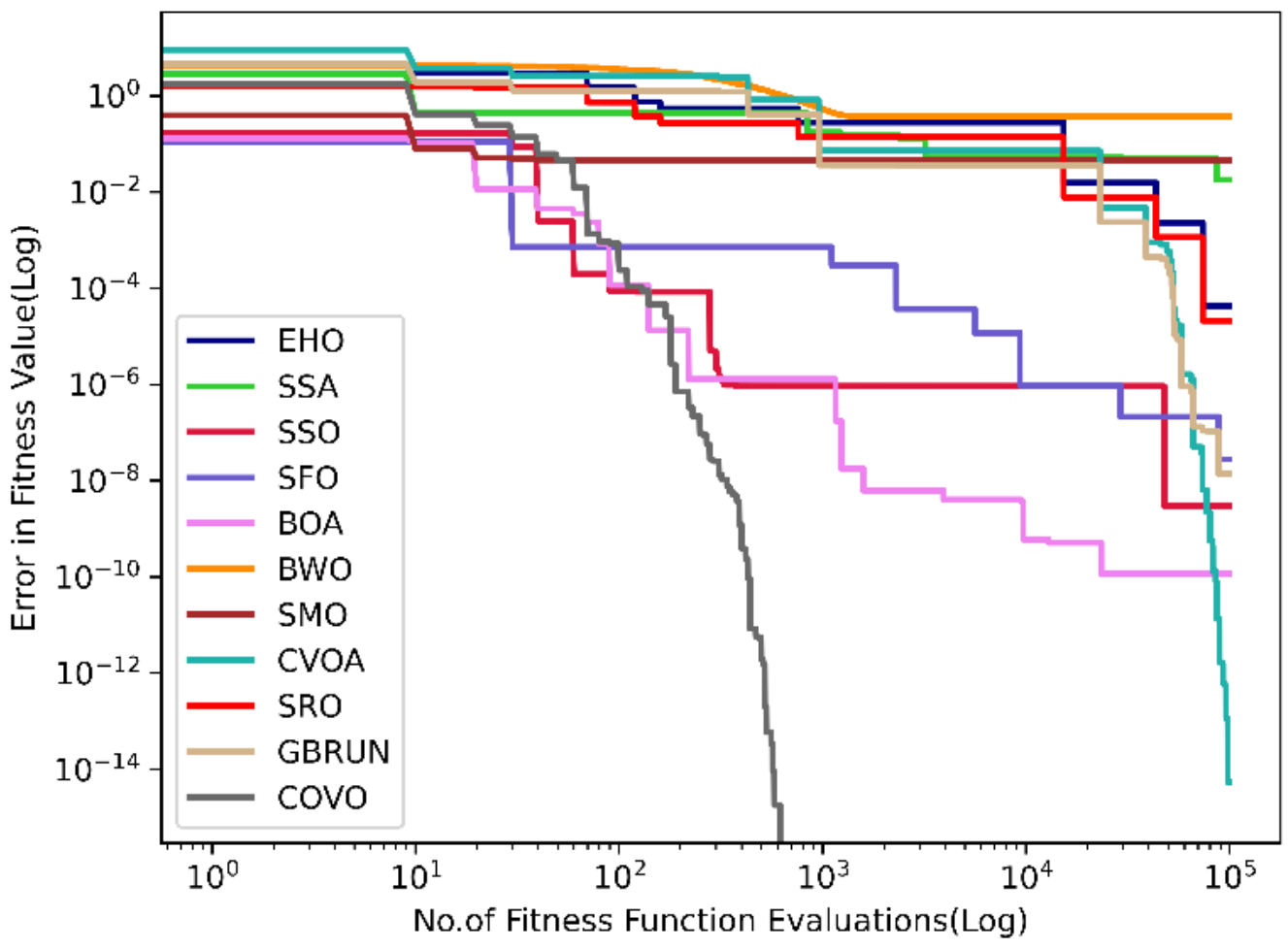

(i)

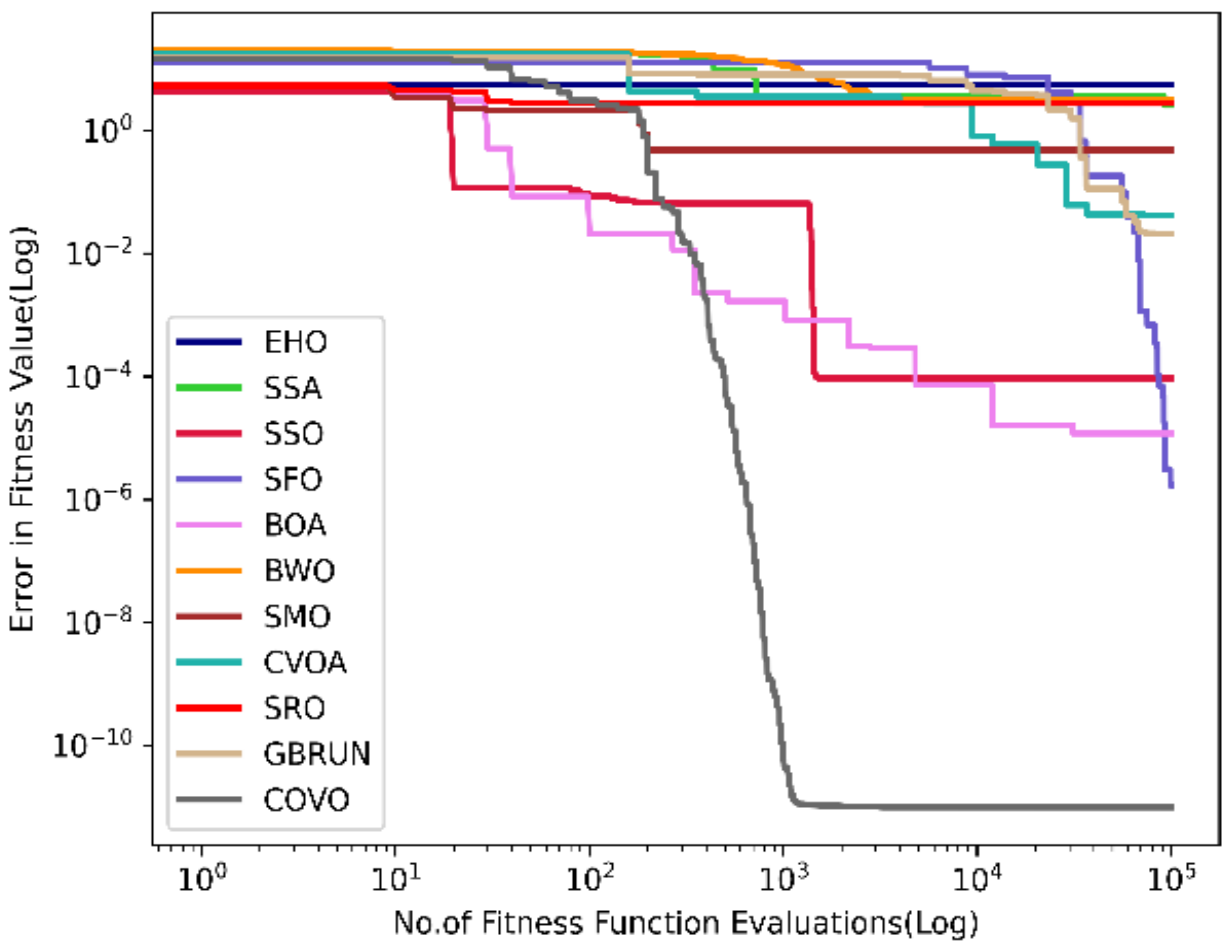

(j)

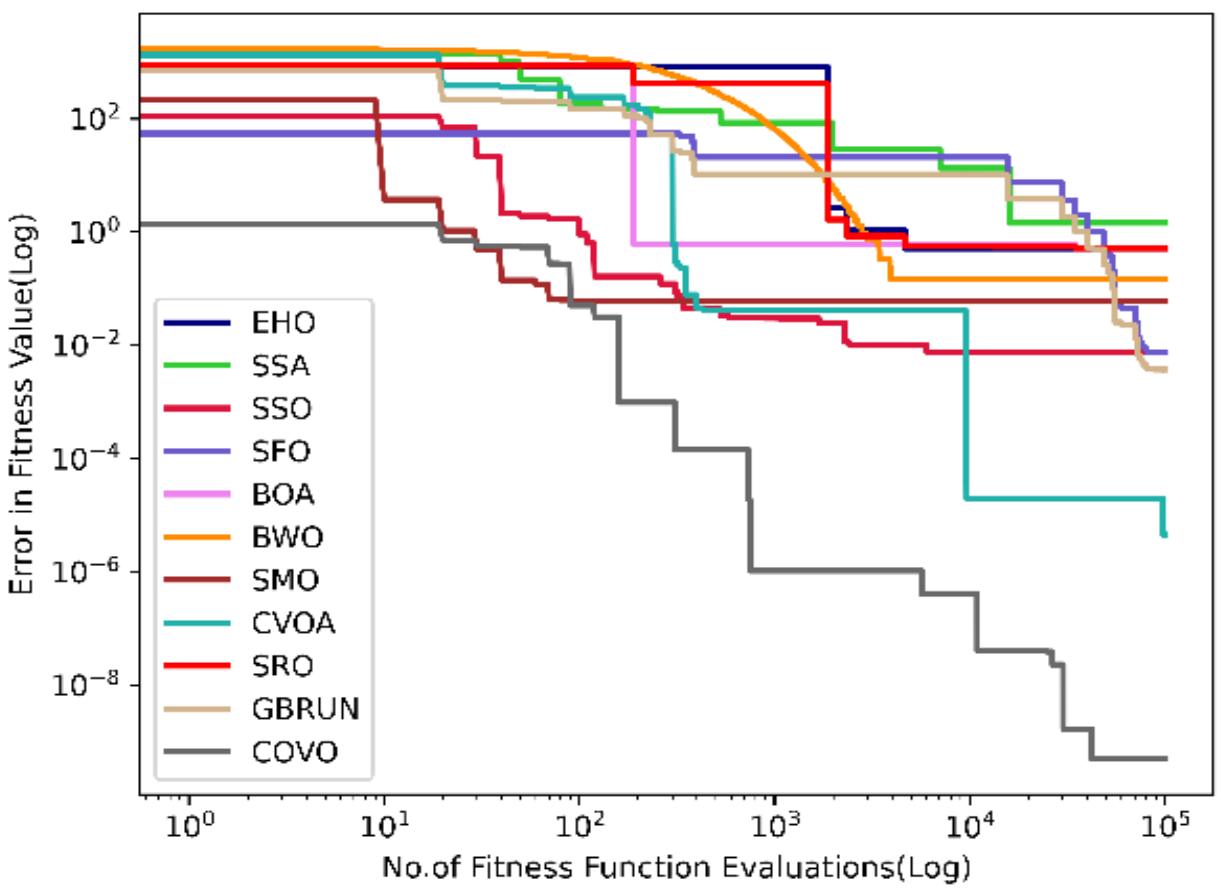

(k)

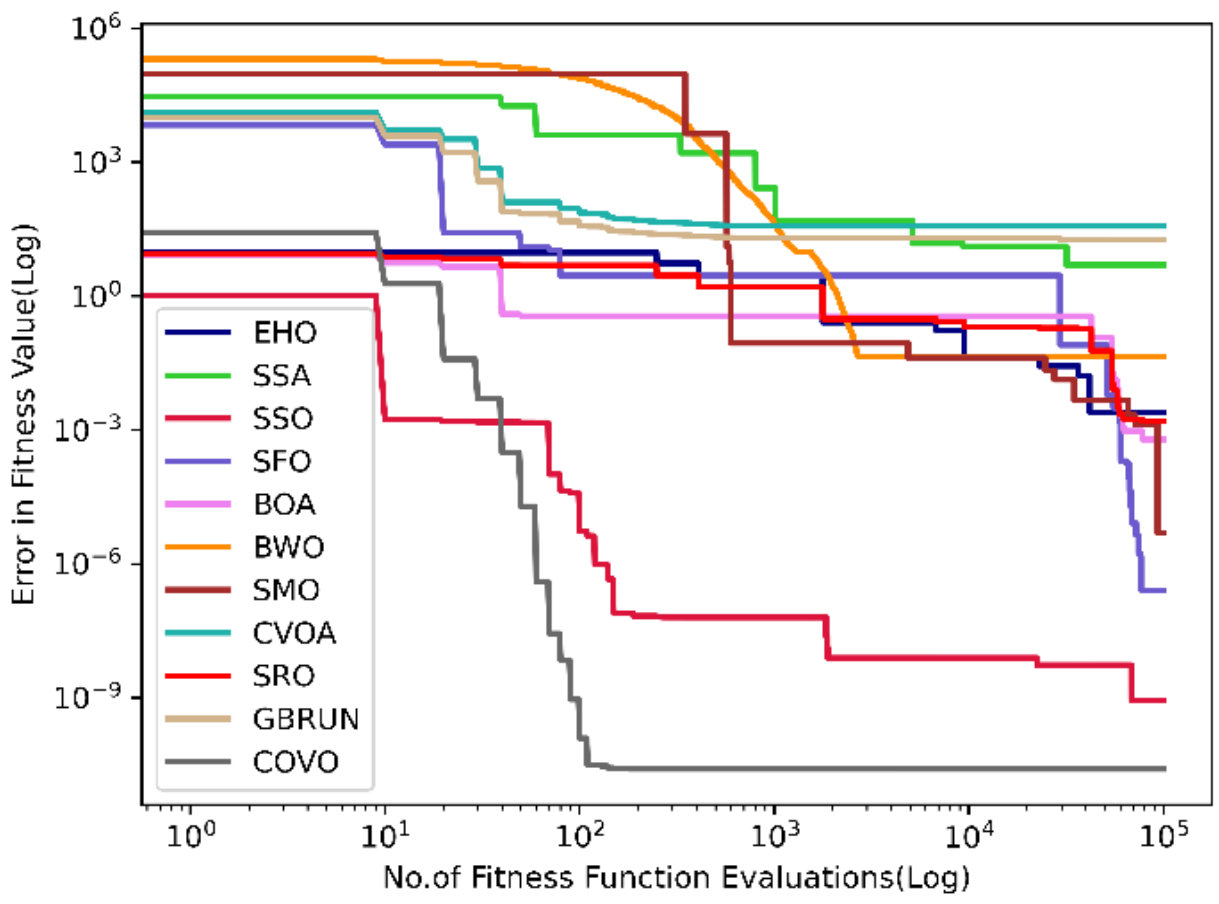

(l)

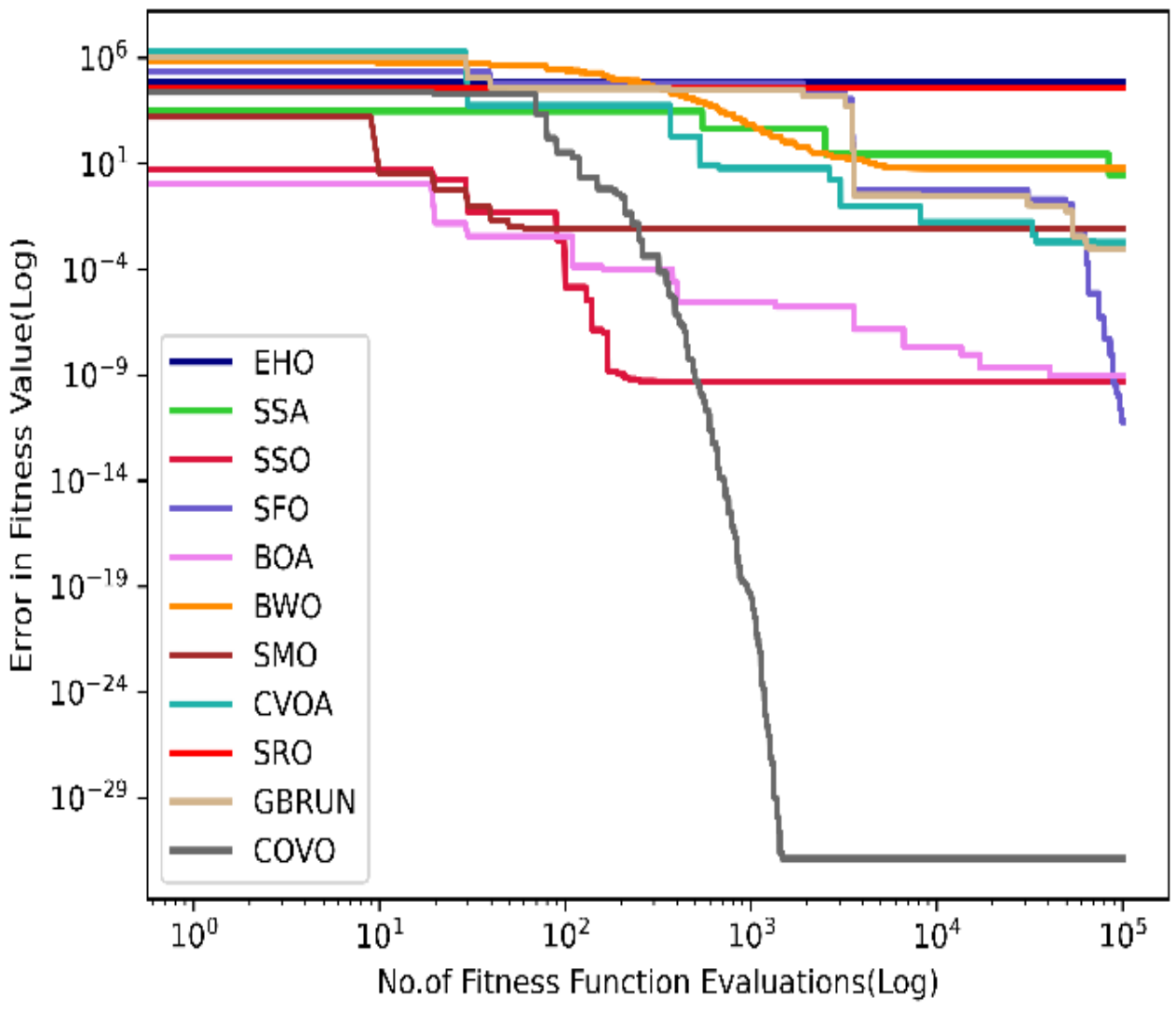

(m)

Fig. 5 Convergence Analysis for (a) $F_1$ (b) $F_2$ (c) $F_3$ (d) $F_4$ (e) $F_5$ (f) $F_6$ (g) $F_7$ (h) $F_8$ (i) $F_9$ (j) $F_{10}$ (k) $F_{11}$ (l) $F_{12}$ (m) $F_{13}$

### A. *Convergence Analysis*

The convergence analysis has been done for the 13 standard benchmark fitness functions $(F_1 - F_4)$ (unimodal) $(F_6 - F_7)$ and $(F_9 - F_{11})$ (multimodal). The convergence of the COVO algorithm is compared with EHO [35], SSO [40], SSA [44], SFO [42], BOA [41], BWO [43], SMO [32], CVOA [47], SRO [53], and GBRUN [54] respectively. The error in fitness value acquired for the 13 standard benchmark functions is shown in Figure 5. It has been observed that the COVO algorithm has performed better than the traditional models in all 13 cases; hence, the COVO algorithm is highly convergent. For function $F_1$, the cost function of COVO is lower than the existing models at the higher count of iterations, and this clearly says that the proposed model can be applied for acquiring global solutions even for complex optimization problems. The fitness function $F_2$ is lower with COVO, even under higher variation in the count of the fitness evaluations. The error in fitness function $F_{13}$ is lower than the existing models like EHO [35], SSO [40], SSA [44], SFO [42], BOA [41], BWO [43], SMO [32], CVOA [47], SRO [53], and GBRUN [54] respectively. The convergence performance of various optimization methods compared to the proposed COVO algorithm shows varying levels of effectiveness. EHO achieves a fitness value of $2.99\times10^{-2}$, indicating moderate convergence, while SSA struggles with a much higher value of $1.80\times10^{2}$. SSO shows a fitness value of $4.69\times10^{-4}$, reflecting reasonable convergence, but still trails behind COVO. SFO demonstrates weak convergence with a value of $3.72\times10^{1}$, and BOA achieves a fitness value of $3.83\times10^{-3}$, which, although better, remains less effective than COVO. BWO's fitness value of 1.99 indicates moderate convergence, whereas SMO performs poorly with a value 4.36. As the number of fitness evaluations increases, algorithms typically refine their search and improve solution quality. COVO's ability to maintain lower error rates throughout this process suggests that it effectively utilizes each evaluation to enhance solution accuracy and avoid degradation in performance. Thus, the COVO algorithm is proven to provide a higher convergence speed to the solutions and can be applied to different applications.

### B. *Statistical Analysis*

The statistical evaluation has been conducted based on 5 significant aspects: mean, median, best, standard deviation, and worst. All these evaluations have been carried out for 13 standard benchmark fitness functions $(F_1 - F_4)$ (unimodal) $(F_6 - F_7)$ and $(F_9 - F_{11})$ (multimodal). COVO's mean values for functions like $F_1$ and $F_2$ are very low, showing that the algorithm consistently performs well on these benchmarks. However, mean values for functions like $F_5$ and $F_7$ are higher, which might suggest less effectiveness compared to other functions. The outcomes observed are manifested in Table IV. With a fitness function error of $1.23\times10^{-18}$, COVO demonstrates exceptional precision in its optimization capability. In contrast, SMO encounters greater difficulty in optimization tasks, with an error of 1.04. SSA, with an error of 10.12, also struggles, highlighting its limitations in achieving the same level of accuracy as COVO in this scenario. The performance in multimodal function optimization and noise removal applications is demonstrated through its low best and median fitness values, consistent mean values, and relatively low standard deviation. COVO's ability to efficiently explore and exploit the search space leads to faster convergence on high-quality solutions. This is evident from the algorithm's superior mean and median fitness values, which indicate that COVO consistently finds better solutions more quickly than its competitors.

TABLE IV STATISTICAL ANALYSIS FOR 13 STANDARD BENCHMARK FUNCTIONS $F_1$ TO $F_{13}$

| **Fitness** | **F1** | **F2** | **F3** | **F4** | **F5** | **F6** | **F7** | **F8** | **F9** | **F10** | **F11** | **F12** | **F13** |
|---|---|---|---|---|---|---|---|---|---|---|---|---|---|
| **Best** | 1.23E-18 | 9.26E-15 | 2.14E-19 | 1.08E-11 | 7.34E-11 | 3.39E-25 | 0.015626 | 3.83E-05 | 0 | 9.89E-12 | 4.93E-10 | 2.50E-11 | 1.35E-32 |
| **Median** | 1.23E-18 | 1.97E-12 | 2.14E-19 | 1.08E-11 | 7.34E-11 | 3.39E-25 | 0.02174 | 3.83E-05 | 0 | 9.89E-12 | 4.93E-10 | 2.50E-11 | 1.35E-32 |
| **Worst** | 3.760311 | 6.88E-05 | 1.07E-12 | 1.646739 | 0.105048 | 0.966721 | 0.16595 | 8.05E-05 | 0.000239 | 3.095663 | 0.050144 | 1.22E-10 | 34.01654 |
| **Mean** | 0.001051 | 1.42E-07 | 1.08E-16 | 0.000517 | 8.06E-05 | 0.000426 | 0.022197 | 3.83E-05 | 7.12E-08 | 0.002303 | 2.48E-05 | 2.50E-11 | 0.006269 |
| **Standard deviation** | 0.051474 | 3.00E-06 | 1.07E-14 | 0.022834 | 0.002401 | 0.016667 | 0.013364 | 4.23E-07 | 3.09E-06 | 0.072481 | 0.000948 | 9.76E-13 | 0.391216 |

### *C. Application of Social Distancing Induced COVO for noise removal*

To portray that the proposed work applies to solving a complex optimization problem, we have validated it with a signal noise removal application. Initially, we collected the ECG signals $S(t)$ and contaminated them with Gaussian noise $N(t)$ as $Y(t) = S(t) * N(t)$. The primary assumption behind the Independent Component Analysis (ICA) problem is that the source signals are independent during the recording period. The contaminated signal $Y(t)$ is subjected to Independent component analysis (ICA) for noise removal. In order to reconstruct the original ECG signals $S(t)$ precisely, the unmixing matrix $W$ of ICA is optimized using the COVO model. In fact, ICA has been applied to diverse applications; the notable one among them is blind source separation. The ECG, which has been infected by Gaussian noise with mean=0, standard deviation=1, and variance=1, is presented for de-noising using COVO and the fitness function. Back projecting the signal from the demixed version to eliminate noise is made using the inverse of the optimized matrix W. The discrepancy between the original and reconstructed ECG is measured using Mean Square Error (MSE). COVO's robust performance in noise removal tasks is attributed to its ability to handle noisy environments effectively. By maintaining a controlled interaction between solutions, COVO reduces the impact of noise on the optimization process and achieves cleaner results.

The residual noise and ECG distortion after the filtering procedure are the primary causes of MSE. The findings are shown in Table V, which shows the MSE between the original signal and the reconstructed signal, which was accomplished using COVO to remove noise. The MSE has also been used to compare the original and contaminated signals. MSEo-r is the MSE between the original and reconstructed ECG signal; MSEo-c is the MSE between the original and Gaussian noise-contaminated signal. Results for MSE between the original and reconstructed ECG demonstrate that COVO improves the demixing matrix for noise reduction. For a given interval, the original ECG, contaminated ECG, and reconstructed ECG are presented in Figure 6.

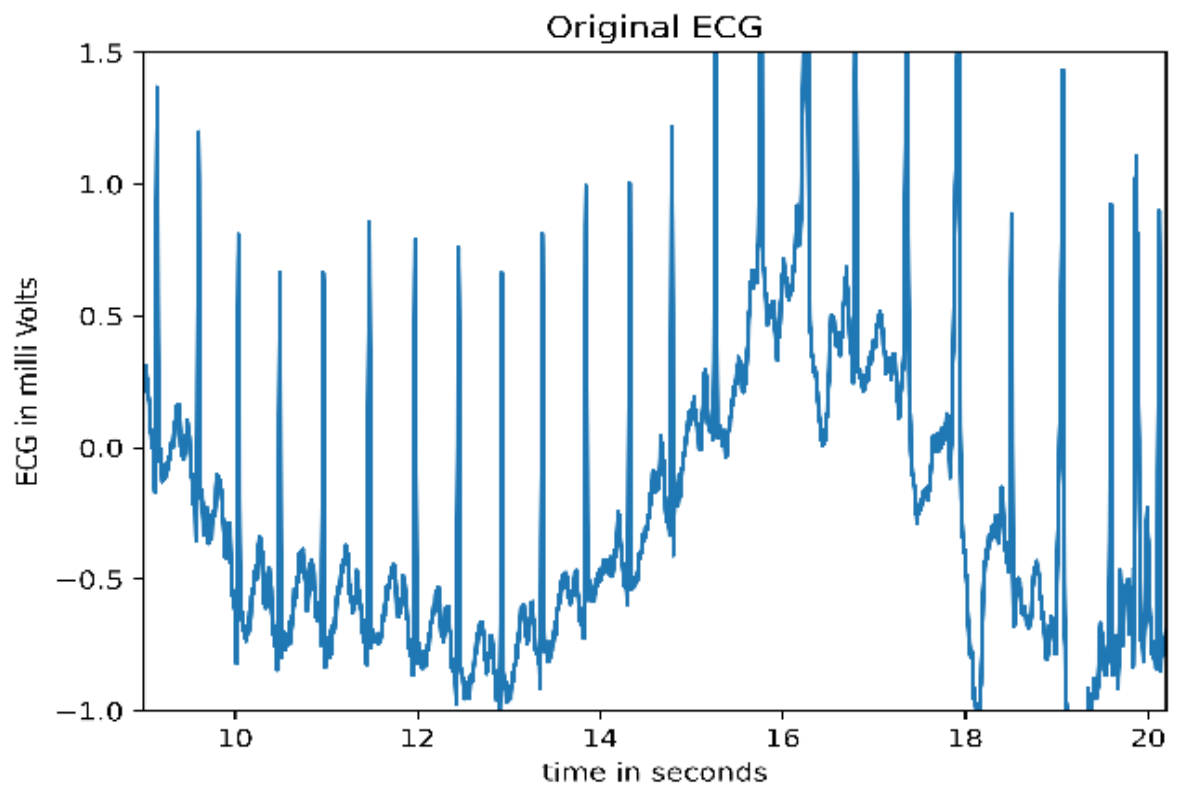

(a)

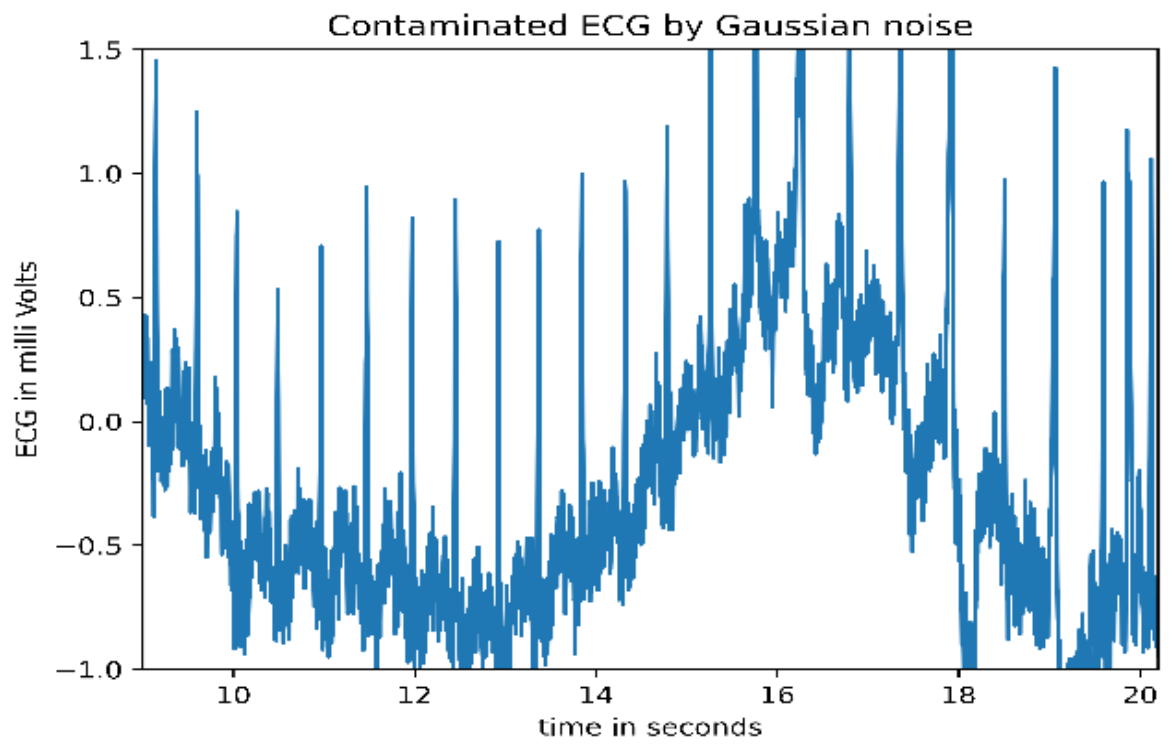

(b)

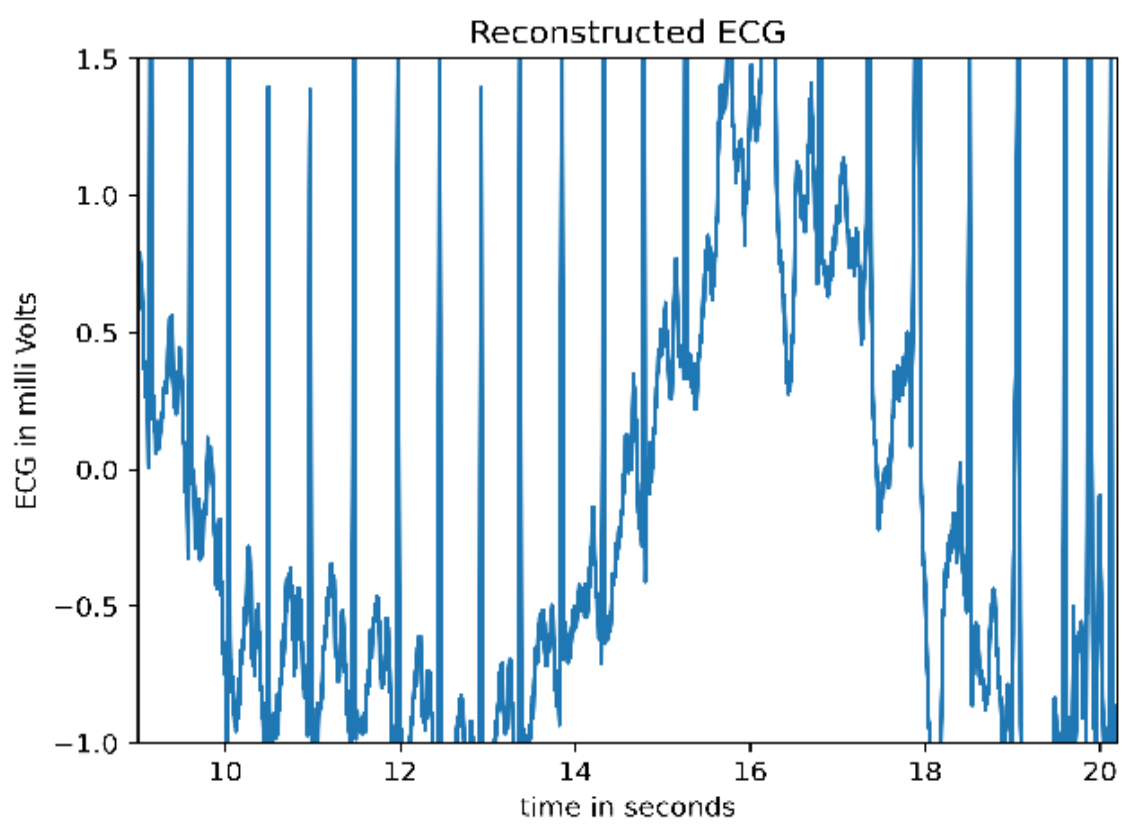

(c)

Fig. 6 (a) ECG Original Signal (b) Contaminated ECG Signal (Gaussian Noise Implied), (c) Reconstructed Signal

TABLE V MSE ACHIEVED BY SOCIAL DISTANCING INDUCED CORONAVIRUS OPTIMIZATION ALGORITHM (COVO)

| Sample Size | MSEo-r | MSEo-c |
|---|---|---|
| **100** | 1.584837 | 0.997821 |
| **101** | 1.532555 | 0.99999 |
| **102** | 0.18787 | 0.99889 |
| **103** | 0.18787 | 0.998061 |

### *D. Computational Time Analysis*

The time metrics indicate the computational time required by each algorithm to solve a set of benchmark problems. Lower time values suggest better performance in terms of computational efficiency. TABLE VI displays the time analysis of the proposed method.

TABLE VI COMPUTATIONAL ANALYSIS

| Methods | Time |
|---|---|
| EHO | 138.7298 |
| SSA | 172.8703 |
| SSO | 135.7377 |
| SFO | 178.9812 |
| BOA | 118.7022 |
| BWO | 189.507 |
| SMO | 132.1517 |
| CVOA | 152.2195 |
| SRO | 149.6909 |
| GBRUN | 179.3319 |
| COVO | 100.5363 |

### *E. Statistical Test Analysis*

The p-value from the Friedman test indicates the probability of observing the test results under the null hypothesis, which states that there are no significant differences between the algorithms. The p-value is above 0.05, suggesting that differences in performance between EHO and other algorithms are not statistically significant. Common significance levels are 0.05 or 0.01. If a p-value is less than the chosen significance level, it indicates a statistically significant difference between algorithms. Table VII and VIII Friedmann and Wilcoxon Test Analysis.

TABLE VII FRIEDMANN TEST

| Methods | P-Value | Static Value |
|---|---|---|
| EHO | 0.541139 | 0.181307 |
| SSA | 0.719659 | 0.218151 |
| SSO | 0.885113 | 0.219914 |
| SFO | 0.531468 | 0.461309 |
| BOA | 0.868234 | 0.297717 |
| BWO | 0.473543 | 0.186363 |
| SMO | 0.876983 | 0.193595 |
| CVOA | 0.848633 | 0.415392 |
| SRO | 0.517019 | 0.129773 |
| GBRUN | 0.505988 | 0.546247 |
| COVO | 0.925938 | 0.104721 |

The Wilcoxon test results indicate that there are no statistically significant differences in the performance metrics of the algorithms tested, as evidenced by the high p-values. This is the lowest p-value among the provided data, but it is still above the common significance level of 0.05. While it is closer to being significant, it does not reach the threshold to be considered statistically significant

TABLE VIII WILCOXON TEST

| Methods | P-Value |
|---|---|
| EHO | 0.533832 |
| SSA | 0.820619 |
| SSO | 0.555022 |
| SFO | 0.727104 |
| BOA | 0.7113 |
| BWO | 0.914671 |
| SMO | 0.309957 |
| CVOA | 0.40239 |
| SRO | 0.27543 |
| GBRUN | 0.241993 |
| COVO | 0.127665 |

### F. *P-Test and T-test*

Table IX provides the analysis of P-Test and T-Test. Most algorithms have relatively high p-values in the P-Test results. For instance, COVO has the highest p-value (0.936619), suggesting that differences in performance metrics between COVO and other algorithms are not significant. Higher T-Test values indicate a larger difference between the means, while lower values suggest smaller differences.

TABLE IX ANALYSIS OF P-TEST AND T-TEST

| Methods | P-Test | T-Test |
|---|---|---|
| EHO | 0.865013 | 1.099513 |
| SSA | 0.707142 | 0.218535 |
| SSO | 0.758872 | 1.77396 |
| SFO | 0.676652 | 0.595475 |
| BOA | 0.466157 | 1.118362 |
| BWO | 0.873233 | 0.334699 |
| SMO | 0.667658 | 1.116085 |
| CVOA | 0.738324 | 1.371728 |
| SRO | 0.776605 | 0.357432 |
| GBRUN | 0.530314 | 1.476108 |
| COVO | 0.936619 | 0.104813 |

## V. CONCLUSION

In this paper, a unique nature-inspired metaheuristic optimization technique for global optimization problems, called the Social Distancing Induced Coronavirus Optimization Algorithm (COVO), is described. The COVO algorithm uses a social distance strategy to try to stop the COVID-19 pandemic from spreading. This study uses 13 benchmark functions to evaluate the proposed COVO with eight cutting-edge metaheuristic approaches. Additionally, the three real-world engineering challenges from IEEE-CEC 2011 are used to validate the obtained optimization. In fact, the outcomes demonstrate that the COVO model works well and will be widely used in the ensuing decades to address a wide range of real-world issues. In addition, with the parameter-free COVO model, parameters self-improve in the future. Moreover, versions that are binary, discrete, and multi-objective are taken into consideration for further investigation.

**Declarations**

**Conflicts of Interest**

The authors have no competing interests to declare that are relevant to the content of this article.

**Authors Contribution Statement**

Om Ramakisan Varma: Conceptualization and model design, formal analysis, manuscript writing, and editing.

Mala Kalra: Validation, manuscript review, editing, and supervision.

**Ethical and Informed Consent for Data Used**

The data used in this study is derived from examples cited in the references, and ethical approval along with informed consent were obtained from all contributing authors.

**Data Availability and Access**

The datasets used or analysed during the current study are available from the corresponding author on reasonable request.

## *References*

[1] H. Emami, "Anti-coronavirus optimization algorithm," *Soft Comput.*, vol. 26, no. 11, pp. 4991–5023, 2022, doi: 10.1007/s00500-022-06903-5.

[2] P. Trojovský and M. Dehghani, "A new optimization algorithm based on mimicking the voting process for leader selection," *PeerJ Comput. Sci.*, vol. 8, pp. 1–40, 2022, doi: 10.7717/peerj-cs.976.

[3] E. H. Houssein, B. E. D. Helmy, D. Oliva, P. Jangir, P. Manoharan A. A. Elngar, and H. Shaban, "An efficient multi-thresholding based COVID-19 CT images segmentation approach using an improved equilibrium optimizer," *Biomed. Signal Process. Control*, vol. 73, no. July 2021, p. 103401, 2022, doi: 10.1016/j.bspc.2021.103401.

[4] D. W. Zingg, M. Nemec, and T. H. Pulliam, "A comparative evaluation of genetic and gradient-based algorithms applied to aerodynamic optimization," *Eur. J. Comput. Mech.*, vol. 17, no. 1–2, pp. 103–126, 2008, doi: 10.3166/REMN.17.103-126.

[5] K. V. Price, R. M. Storn, and J. Lampinen, *Differential Evolution-A Practical Approach to Global Optimization*. 2005. doi: 10.1007/3-540-31306-0_12.

[6] X. S. Yang, "Cuckoo Search and Firefly Algorithm: Overview and Analysis," *Stud. Comput. Intell.*, vol. 585, pp. 1–26, 2014, doi: 10.1007/978-3-319-02141-6.

[7] X. S. Yang, S. F. Chien, and T. O. Ting, "Bio-Inspired Computation and Optimization: An Overview", *Bio-Inspired Computation in Telecommunications.*, pp. 1-21, 2015. doi: 10.1016/B978-0-12-801538-4.00001-X.

[8] A. Afshar, O. Bozorg Haddad, M. A. Mariño, and B. J. Adams, "Honey-bee mating optimization (HBMO) algorithm for optimal reservoir operation," *J. Franklin Inst.*, vol. 344, no. 5, pp. 452–462, 2007, doi: 10.1016/j.jfranklin.2006.06.001.

[9] S. Mirjalili, S. M. Mirjalili, and A. Lewis, "Grey Wolf Optimizer," *Adv. Eng. Softw.*, vol. 69, pp. 46–61, 2014, doi: 10.1016/j.advengsoft.2013.12.007.

[10] M. A. Al-Betar, Z. A. A. Alyasseri, M. A. Awadallah, and I. Abu Doush, *Coronavirus herd immunity optimizer (CHIO), Neural Comput Appl.*, vol. 33, no. 10. pp. 5011-5042, 2021. doi: 10.1007/s00521-020-05296-6.

[11] S. Mirjalili and A. Lewis, "The Whale Optimization Algorithm," *Adv. Eng. Softw.*, vol. 95, pp. 51–67, 2016, doi: 10.1016/j.advengsoft.2016.01.008.

[12] F. Fausto, A. Reyna-Orta, E. Cuevas, Á. G. Andrade, and M. Perez-Cisneros, "From ants to whales: metaheuristics for all tastes," *Artif Intell Rev*, vol. 53, no. 1. pp. 753-810, 2020. doi: 10.1007/s10462-018-09676-2.

[13] J.H. Holland., "Genetic Algorithms and Adaptation," *Adaptive Control of Ill-Defined Systems. NATO Conference Series*., vol. 16, pp. 317–333, 1984. doi.org/10.1007/978-1-4684-8941-5_21

[14] J. Shapiro, "Genetic algorithms in machine learning," *Machine Learning and Its Applications,* vol. 2049, pp. 146–168, 2001, doi: 10.1007/3-540-44673-7_7.

[15] J. Kennedy and E. Russell, "Particle Swarm Optimisation," *in Proceedings of ICNN'95 - International Conference on Neural Networks*, pp. 1942–1948. 1995, doi: 10.1007/978-3-030-61111-8_2.

[16] M. Dorigo and G. Di Caro, "Ant colony optimization: A new meta-heuristic," *Proc. 1999 Congr. Evol. Comput. CEC 1999*, vol. 2, pp. 1470–1477, 1999, doi: 10.1109/CEC.1999.782657.

[17] S. Kirkpatrick, C. D. Gelatt, and M. P. Vecchi, "Optimization by Simulated Annealing," *Sci. 220(4598)*, vol. 220, no. 4598, pp. 671–

680, 1983, doi: 10.1126/science.220.4598.671
[18] Z. W. Geem, J. H. Kim, and G. V. Loganathan, "A New Heuristic Optimization Algorithm: Harmony Search," *Simulation*, vol. 76, no. 2, pp. 60–68, 2001, doi: 10.1201/b18469-4.
[19] X. S. Yang, "Firefly algorithms for multimodal optimization," *Lect. Notes Comput. Sci. (including Subser. Lect. Notes Artif. Intell. Lect. Notes Bioinformatics)*, vol. 5792, pp. 169–178, 2009, doi: 10.1007/978-3-642-04944-6_14.
[20] D. H. Wolpert and Macready William G., "No Free Lunch Theorems for Optimization," *IEEE Trans. Evol. Comput.*, vol. 1, no. 1, pp. 67–81, 1997. doi: 10.1109/4235.585893
[21] M. A. Al-Betar, "β-Hill climbing: an exploratory local search," *Neural Comput. Appl.*, vol. 28, no. s1, pp. 153–168, 2017, doi: 10.1007/s00521-016-2328-2.
[22] E. Hosseini, K. Z. Ghafoor, A. S. Sadiq, M. Guizani, and A. Emrouznejad, "COVID-19 Optimizer Algorithm, Modeling and Controlling of Coronavirus Distribution Process," *IEEE J. Biomed. Heal. Informatics*, vol. 24, no. 10, pp. 2765–2775, 2020, doi: 10.1109/JBHI.2020.3012487.
[23] K. Golalipour, I. F. Davoudkhani, S. Nasri, A. Naderipour, S. Mirjalili, A. Y. Abdelaziz., "The corona virus search optimizer for solving global and engineering optimization problems," *Alexandria Eng. J.*, vol. 78, no. July, pp. 614–642, 2023, doi: 10.1016/j.aej.2023.07.066.
[24] C. P. Robert, "The Metropolis– Hastings Algorithm," *Monte Carlo Statistical Methods*, pp. 231–283, 2015, doi.org/10.1007/978-1-4757-3071-5_6
[25] M. Dorigo and D. C. Gianni, "Ant Colony Optimization: A New Meta-Heuristic," *Proceedings of the 1999 congress on evolutionary computation-CEC99 (Cat. No. 99TH8406)*. pp. 1470–1477, 1992, doi: 10.1109/CEC.1999.782657
[26] A. H. Gandomi, X. S. Yang, and A. H. Alavi, "Cuckoo search algorithm: A metaheuristic approach to solve structural optimization problems," *Eng. Comput.*, vol. 29, no. 1, pp. 17–35, 2013, doi: 10.1007/s00366-011-0241-y.
[27] F. Glover and M. Laguna, "Tabu Search," *Handbook of Combinatorial Optimization*, vol. 3, pp. 621–757, 1998, doi.org/10.1007/978-1-4613-0303-9_33
[28] J. D. Farmer, N. H. Packard, and A. S. Perelson, "The Immune System, Adaptation, and Machine Learning," *Phys. D*, vol.22 no. 1-3 pp. 187–204, 1986, doi:dx.doi.org/10.1016/0167-2789(86)90240-X
[29] X. S. Yang, "Engineering optimizations via nature-inspired virtual bee algorithms," *Lect. Notes Comput. Sci.*, vol. 3562, no. 2, pp. 317–323, 2005, doi: 10.1007/11499305_33.
[30] D. Karaboga, "An Idea Based on Honey Bee Swarm for Numerical Optimization," *Tech. report-tr06, Erciyes Univ. Eng. Fac. Comput. Eng. Dep.*, vol.200, pp. 1–10, 2005, doi: doi=10.1.1.714.4934.
[31] X. S. Yang and A. H. Gandomi, "Bat algorithm: A novel approach for global engineering optimization," *Eng. Comput. (Swansea, Wales)*, vol. 29, no. 5, pp. 464–483, 2012, doi: 10.1108/02644401211235834.
[32] J. C. Bansal, H. Sharma, S. S. Jadon, and M. Clerc, "Spider Monkey Optimization algorithm for numerical optimization," *Memetic Comput.*, vol. 6, no. 1, pp. 31–47, 2014, doi: 10.1007/s12293-013-0128-0.
[33] A. Sharma, A. Sharma, B. K. Panigrahi, D. Kiran, and R. Kumar, "Ageist Spider Monkey Optimization algorithm," *Swarm Evol. Comput.*,vol. 28, pp. 58–77, 2016, doi: 10.1016/j.swevo.2016.01.002.
[34] Z. Wang, J. Mumtaz, L. Zhang, and L. Yue, "Application of an improved spider monkey optimization algorithm for component assignment problem in PCB assembly," *Procedia CIRP*, vol. 83, pp. 266–271, 2019, doi: 10.1016/j.procir.2019.04.075.
[35] G. G. Wang, S. Deb, and L. D. S. Coelho, "Elephant Herding Optimization," *Proc. - 2015 3rd Int. Symp. Comput. Bus. Intell. ISCBI 2015*, pp. 1–5, 2016, doi: 10.1109/ISCBI.2015.8.
[36] Y. C. Liang and J. R. Cuevas Juarez, "A novel metaheuristic for continuous optimization problems: Virus optimization algorithm," *Eng. Optim.*, vol. 48, no. 1, pp. 73–93, 2016, doi: 10.1080/0305215X.2014.994868.
[37] Y.C. Liang, and J. R. Cuevas Juarez, "A self-adaptive virus optimization algorithm for continuous optimization problems", *Soft Comput*, vol. 24, pp 13147–13166, 2020, doi: 10.1007/s00500-020-04730-0.
[38] A. Salehan and A. Deldari, "Corona virus optimization (CVO): a novel optimization algorithm inspired from the Corona virus pandemic," *J. Supercomput.*, vol. 78, no. 4, pp. 5712–5743, 2022, doi: 10.1007/s11227-021-04100-z.
[39] M. A. Shamseldin, "Optimal coronavirus optimization algorithm based pid controller for high performance brushless dc motor," *Algorithms*, vol. 14, no. 7, pp. 1-17, 2021, doi: 10.3390/a14070193.
[40] S. Mirjalili, A. H. Gandomi, S. Z. Mirjalili, S. Saremi, H. Faris, and S. M. Mirjalili, "Salp Swarm Algorithm: A bio-inspired optimizer for engineering design problems," *Adv. Eng. Softw.*, vol. 114, pp. 163–191, 2017, doi: 10.1016/j.advengsoft.2017.07.002.
[41] S. Arora and S. Singh, "Butterfly optimization algorithm: a novel approach for global optimization," *Soft Comput.*, vol. 23, no. 3, pp. 715–734, 2019, doi: 10.1007/s00500-018-3102-4.
[42] S. Shadravan, H. R. Naji, and V. K. Bardsiri, "The Sailfish Optimizer: A novel nature-inspired metaheuristic algorithm for solving constrained engineering optimization problems," *Eng. Appl. Artif. Intell.*, vol. 80, no. 1, pp. 20–34, 2019, doi: 10.1016/j.engappai.2019.01.001.
[43] V. Hayyolalam and A. A. Pourhaji Kazem, "Black Widow Optimization Algorithm: A novel meta-heuristic approach for solving engineering optimization problems," *Eng. Appl. Artif. Intell.*, vol. 87, no. 1, pp. 1-28, 2020, doi: 10.1016/j.engappai.2019.103249.
[44] J. Xue and B. Shen, "A novel swarm intelligence optimization approach: sparrow search algorithm," *Syst. Sci. Control Eng.*, vol. 8, no. 1, pp. 22–34, 2020, doi: 10.1080/21642583.2019.1708830.
[45] M. Hubálovská, Š. Hubálovský, and P. Trojovský, "Botox Optimization Algorithm: A New Human-Based Metaheuristic Algorithm for Solving Optimization Problems," *Biomimetics*, vol. 9, no. 3, pp. 1-41, 2024, doi: 10.3390/biomimetics9030137.
[46] O. Al-Baik, S. Alomari, O. Alssayed, S. Gochhait, I. Leonova,U. Dutta, O. P. Malik, Z. Montazeri, M. Dehghani, "Pufferfish Optimization Algorithm: A New Bio-Inspired Metaheuristic Algorithm for Solving Optimization Problems," *Biomimetics*, vol. 9, no. 2, pp. 1-54, 2024, doi: 10.3390/biomimetics9020065.
[47] F. Martínez-Álvarez, G. Asencio-Cortés, J.F.Torres, D. Gutiérrez-Avilés, L. Melgar-García, R. Pérez-Chacón, C. Rubio-Escudero, J.C. Riquelme, and A. Troncoso, "Coronavirus Optimization Algorithm: A Bioinspired Metaheuristic Based on the COVID-19 Propagation Model," *Big Data*, vol. 8, no. 4, pp. 308–322, 2020, doi: 10.1089/big.2020.0051.
[48] S. Safiullah, A. Rahman, S. A. Lone, S. M. S. Hussain, and T. S. Ustun, "Novel COVID-19 Based Optimization Algorithm (C-19BOA) for Performance Improvement of Power Systems," *Sustain.*, vol. 14, no. 21, pp. 1–27, 2022, doi: 10.3390/su142114287.
[49] A. M. Khalid, H. M. Hamza, S. Mirjalili, and K. M. Hosny, "MOCOVIDOA: a novel multi-objective coronavirus disease optimization algorithm for solving multi-objective optimization problems," *Neural Comput. Appl.*, vol. 35, no. 23, pp. 17319–17347, 2023, doi: 10.1007/s00521-023-08587-w.
[50] V. A. Akash Saxena, Siddharth Singh Chouhan , Rabia Musheer Aziz, "A comprehensive evaluation of Marine predator chaotic algorithm for feature selection of COVID-19," *Evol. Syst.*, vol. 1, no. 1, pp. 1–12, 2024, doi: doi.org/10.1007/s12530-023-09557-2.
[51] H. Selim, A. Y. Haikal, L. M. Labib, and M. M. Saafan, "MCHIAO: a modified coronavirus herd immunity-Aquila optimization algorithm based on chaotic behavior for solving engineering problems," *Neural Comput. & Applic.*, vol. 1, no .4, pp. 1-85, 2024. doi: 10.1007/s00521-024-09533-0.
[52] Y. Yuan, Q. Shen, S. Wang, J. Ren, D. Yang, Q. Yang, J. Fan, and Mu, X, "Coronavirus Mask Protection Algorithm: A New Bio-inspired Optimization Algorithm and Its Applications," *J. Bionic Eng.*, vol. 20, no. 4, pp.1747-1765, 2023, doi: 10.1007/s42235-023-00359-5.
[53] S. C. Chu, T. T. Wang, A. R. Yildiz, and J. S. Pan, "Ship Rescue Optimization: A New Metaheuristic Algorithm for Solving Engineering Problems," J. Internet Technol., vol. 25, no. 1, pp. 61–78, 2024, doi: 10.53106/160792642024012501006.
[54] Z. C. Dou, S. C. Chu, Z. Zhuang, A. R. Yildiz, and J. S. Pan, "GBRUN: A Gradient Search-based Binary Runge Kutta Optimizer for Feature Selection," J. Internet Technol., vol. 25, no. 3, pp. 341–353, 2024, doi: 10.53106/160792642024052503001.
[55] C. Shorten, T. M. Khoshgoftaar, and B. Furht, "Deep Learning

applications for COVID-19,” *J. Big Data*, vol. 8, no. 1, pp.1-54, 2021, doi: 10.1186/s40537-020-00392-9.

[56] “COVID Live - Coronavirus Statistics - Worldometer.” https://www.worldometers.info/coronavirus/ (accessed May 07, 2024).

[57] “Coronavirus vs. SARS: How Do They Differ?” https://www.healthline.com/health/coronavirus-vs-sars#covid-19-vs-sars (accessed May 07, 2024).

[58] C. Sun and Z. Zhai, “The efficacy of social distance and ventilation effectiveness in preventing COVID-19 transmission,” *Sustain. Cities Soc.*, vol. 62, no. 7, pp. 1-10, 2020, doi: 10.1016/j.scs.2020.102390.

[59] “Coronavirus Graphs: Worldwide Cases and Deaths - Worldometer.”https://www.worldometers.info/coronavirus/worldwide-graphs/ (accessed May 07, 2024).

[60] Y. Liu, A. A. Gayle, A. Wilder-Smith, and J. Rocklöv, “The reproductive number of COVID-19 is higher compared to SARS coronavirus,” *J. Travel Med.*, vol. 27, no. 2, pp. 1–4, 2020, doi: 10.1093/jtm/taaa021.

[61] S. Arora and S. Singh, “An improved butterfly optimization algorithm with chaos,” *J. Intell. Fuzzy Syst.*, vol. 32, no. 1, pp. 1079–1088, 2017, doi: 10.3233/JIFS-16798.

[62] A. A. Ewees, M. Abd Elaziz, and E. H. Houssein, “Improved grasshopper optimization algorithm using opposition-based learning,” Expert Syst. Appl., vol. 112, pp. 156–172, 2018, doi: 10.1016/j.eswa.2018.06.023.